\documentclass[traditabstract,onecolumn]{aa}
\usepackage{amsmath}
\usepackage{mathbbol}
\usepackage{graphicx,subfigure}
\usepackage{natbib}
\usepackage[ngerman,american]{babel}
\usepackage{txfonts}

\newcommand{\magnet}{-\frac{1}{2}\textbf{B}^{\dagger}\textbf{M}^{-1}\textbf{B}}
\newcommand{\vfactor}{\frac{1}{\sqrt{\vert 2\pi M} \vert} }
\newcommand{\fourierI}{i(\textbf{k}_{\perp} \textbf{x}_{\perp}-\textbf{k}_{\perp}' \textbf{x}_{\perp}')}
\newcommand{\Ji}{\tilde{\textbf{J}}}

\newcommand{\source}{\left[\frac{1}{2} \textbf{J}^{\dagger}\textbf{M}\textbf{J}\right]}

\newcommand{\half}{\frac{1}{2}}
\newcommand{\jm}[2]{\half (\textbf{J}^{\dagger}\textbf{M})_{#1}(\textbf{#2})}
\newcommand{\mj}[2]{\half (\textbf{M}\textbf{J})_{#1}(\textbf{#2})}
\newcommand{\fe}[1]{\textbf{#1}}
\newcommand{\mm}[4]{\half M_{#1}(\fe{#2},\fe{#3})+\half M_{#4}(\fe{#3},\fe{#2})}

\begin{document}

\title{Imprints of magnetic power and helicity spectra on radio polarimetry statistics}

\author{H. Junklewitz\inst{\ref{Max}}, T. A. En\ss lin\inst{\ref{Max}}}

\institute{Max-Planck-Institut f\"ur Astrophysik, Karl-Schwarzschildstr.1, 85741 Garching,Deutschland}\label{Max}

\date{Received <date> / Accepted <date>}

\abstract{Statistical properties of turbulent magnetic fields in radio-synchrotron sources should imprint on the statistics of polarimetric observables. In search of these imprints, we calculate correlation and cross-correlation functions from a set of observables containing the total intensity $I$, the polarized intensity $P$ and the Faraday depth $\phi$. The correlation functions are evaluated for all combinations of observables up to fourth order in the magnetic field $\textbf{B}$. We derive these as far as possible analytically and from first principles only using some basic assumptions such as Gaussian statistics of the underlying magnetic field in the observed region and statistical homogeneity. We further assume some simplifications to reduce the complexity of the calculations, as for a start we were interested in a proof of concept. Using this statistical approach, we show that it is in principle possible to gain information about the helical part of the magnetic power spectrum, namely via the correlation functions $\langle P(\textbf{k}_{\perp}) \phi(\textbf{k}'_{\perp})\phi(\textbf{k}''_{\perp})\rangle_{\textbf{B}}$ and $\langle I(\textbf{k}_{\perp}) \phi(\textbf{k}'_{\perp})\phi(\textbf{k}''_{\perp})\rangle_{\textbf{B}}$. Using this insight, we construct an easy-to-use test for helicity, called {\it LITMUS} (\fe{L}ocal \fe{I}nference \fe{T}est for \fe{M}agnetic fields which \fe{U}ncovers helice\fe{S}) which gives a spectrally integrated measure of helicity. For now, all calculations are given in a Faraday-free case, but set up in a way so that Faraday rotational effects could be included later on.}

\keywords{Magnetic fields -- Methods: data analysis -- Methods: statistical}

\titlerunning{Imprints of magnetic and helicity spectra}

\authorrunning{Henrik Junklewitz \& Torsten En\ss lin}

\maketitle

\section{Introduction}
	Magnetic fields are observed in almost all astronomical objects, they permeate planets and stars as well as galaxies and clusters. Most, if not all, of the interstellar and intergalactic plasma appears to be magnetized and the magnetic fields contribute significantly to physical processes. Examples include the formation of stars \citep{2009RMxAC..36..128P}, the anisotropy of transport processes (thermal conduction or plasma resistivity, see e.g. \citet{2001ApJ...562L.129N}), the angular momentum transport in accretion discs or the propagation of cosmic ray populations \citep{2007ARNPS..57..285S}.
	
	Although magnetic fields are ubiquitous in the cosmos, we often cannot treat them properly in astrophysical situations due to the lack of knowledge of their properties. Cosmic magnetic fields are difficult to observe and their distribution, evolution and origins are far from being perfectly understood. We have three main sources of information: the Zeeman effect, synchrotron radiation and Faraday rotation. The Zeeman effect is extremely difficult to detect, because other line shifting effects, such as thermal Doppler-broadening, are usually stronger. We obtain a great deal of information from synchrotron radiation but only regarding the magnetic field component perpendicular to the line of sight. In order to get a picture of the 3D magnetic field, one needs another source of information. This leads us to Faraday rotation, the change of the polarisation-plane of long wavelength radiation due to a magnetic field along the line of sight. Faraday rotation provides a powerful tool, but is also difficult to observe, to evaluate, and to interpret due to the involved line of sight projection. This projection is one of the main obstacles to understand the 3D properties of magnetic fields.
	
	One important property of cosmic magnetic fields that we do not know much about is magnetic helicity. It is defined as the integral
	\begin{align}
	&H=\int_{V} \fe{A} \cdot \fe{B} \ dx^3 \label{hel}
	\end{align}
	over a Volume $V$ with surface $\partial V$ on which $\fe{n} \cdot \fe{B}=0$; where $\fe{A}$ refers to the  vector potential from electrodynamics with $\fe{B}=\nabla \times \fe{A}$. Helicity is a measure for the ``spiral quality'' of a magnetic field. It quantifies how much the magnetic field lines are sheared and twisted and counts the number of spirals the field lines exhibit within a given volume. Particularly turbulent magnetic fields should show considerable helicity. The relevance of helicity has increased since its inclusion as an essential element in the magnetic dynamo theory, which tries to explain the sustainement of magnetic fields on large scales over cosmic timescales  \citep[see][]{2002BASI...30..715S,2005PhR...417....1B,2005AN....326..400B}. The possible operation of a large scale dynamo for instance is directly connected to the generation of helicity in turbulent environments \citep[see][]{2006A&A...448L..33S,2009PPCF...51l4043B,2007PPCF...49..447S}, a process which, until now, could not be verified through observation. 

	This study is particularly concerned with the question of how to extract knowledge regarding turbulent magnetic helicity spectra from the statistical information found in radio-observational data involving polarisation and Faraday rotation measurements. It was highly motivated by the studies of \citet{2010JETPL..90..637V}.

	Information on magnetic fields can be imprinted onto radio data by two of the processes already mentioned above: \textit{Synchrotron emission} and \textit{Faraday rotation}. Yet information is not only contained in their mean values but also in higher order correlation and cross-correlation functions.
	
	We therefore investigate a set of suitable radio observables for their cross-correlations, to see how these are connected to the statistical properties of the magnetic fields to be examined. This idea goes back to previous works by \citet{1982ApJ...261..310S,1983ApJ...271L..49S,1989AJ.....98..244E,1989AJ.....98..256E,2003A&A...401..835E,2006PhRvD..73f3507K,2009MNRAS.398.1970W}.
	The set of radio observables we investigate contains the total intensity $I(\fe{x})$, the polarised intensity $P(\fe{x})$ and the Faraday depth $\phi(\fe{x})$. We work out all correlation functions between them in a general framework. We restrict ourselves to fourth order in the magnetic field strength and as far as possible we do all calculations analytically. The aim is to find a direct relation to statistical properties of the magnetic fields, such as their power spectra. 
	
\begin{figure}[!h]
	\centering
	\includegraphics[width=0.8\textwidth]{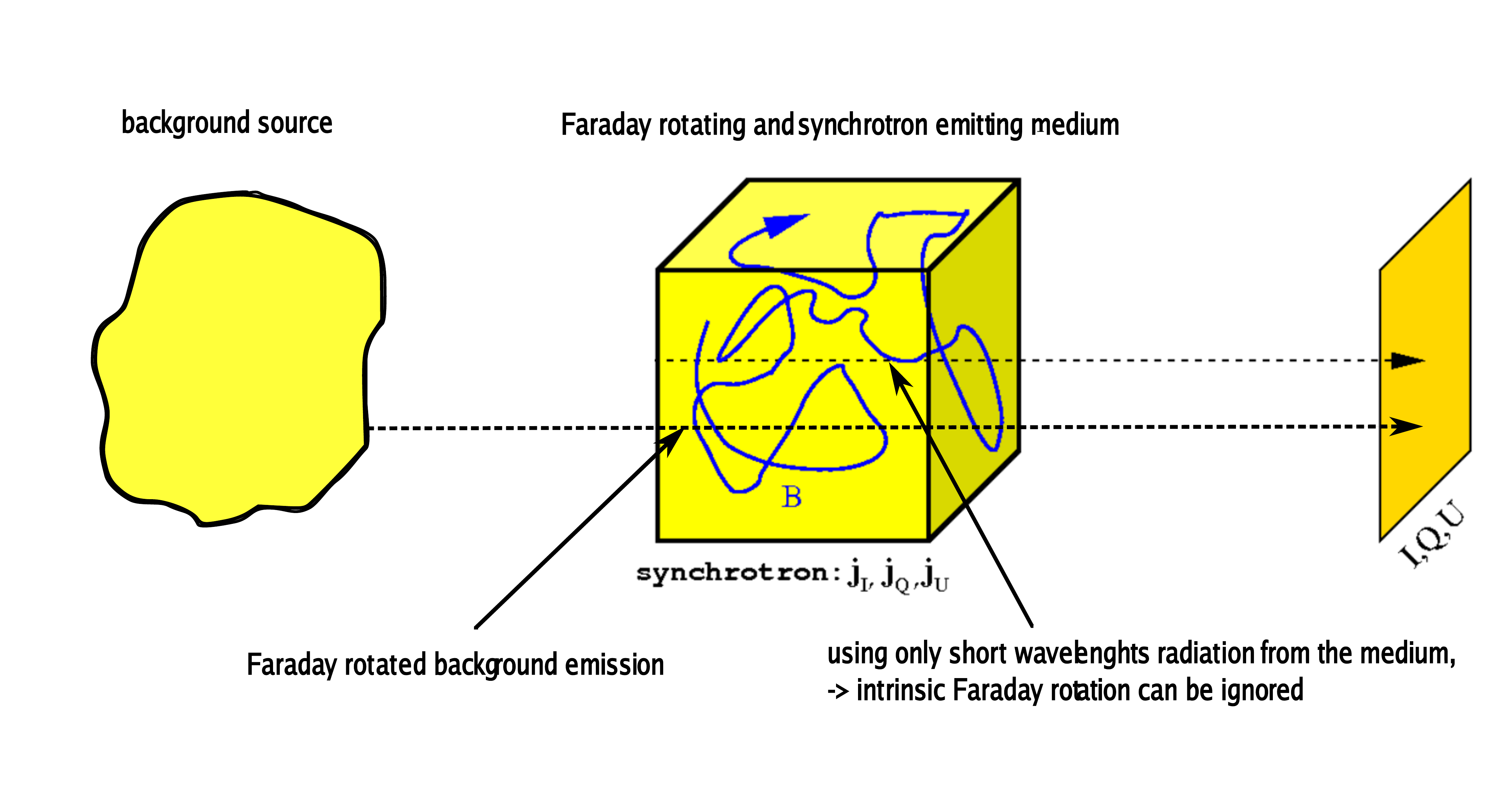}
	\caption{Schematic of an observational situation for which our set of correlation functions is suitable (modified picture taken from \citet{2009MNRAS.398.1970W}).}
	\label{obsreg}
	\end{figure}

	The intensity $I(\fe{x})$ and the polarised intensity $P(\fe{x})$ are connected to the synchrotron emission within a magnetized volume. We assume them to be taken at sufficiently high frequencies and, therefore, free of Faraday rotation. The Faraday depth $\phi(\fe{x})$ is measured via the Faraday rotation of a polarized background source at a different frequency seen through the same volume. The observational situation is visualized in Fig. (\ref{obsreg}).

	With regard to these observable quantities, we can successfully establish all correlation functions in the form of analytical relations to the magnetic field power and helicity spectra implementing Gaussian field statistics for simplicity. The result here is to prove that it is possible, in principle, to gain information not only in respect of the total but also regarding the helical part of the magnetic power spectrum, namely via $\langle P(\textbf{k}_{\perp}) \phi(\textbf{k}'_{\perp})\phi(\textbf{k}''_{\perp})\rangle_{\textbf{B}}$ and $\langle I(\textbf{k}_{\perp}) \phi(\textbf{k}'_{\perp})\phi(\textbf{k}''_{\perp})\rangle_{\textbf{B}}$.

	Gaussian magnetic fields statistics is not what numerical simulations of MHD turbulence find \citep[see][]{2009MNRAS.398.1970W}. However, they are the starting point of any analysis of high-order correlation functions. In case all we know statistically about the fields is their two point correlation, the only assumption which expresses solely this knowledge is a Gaussian with the correlation tensor being the covariance matrix. Any other distribution function would contain more information in a Shannon-Boltzmann sense. 

	If additional information on higher order statistics is available, this could be incorporated via perturbative methods. These would expand around the Gaussian case, which has therefore to be worked out first, as we do in this work.
	
	Based on our results we further present the \textit{LITMUS} test (\fe{L}ocal \fe{I}nference \fe{T}est for \fe{M}agnetic fields which \fe{U}ncovers helice\fe{S}), a first simple procedure to probe data for helicity. An analysis of real and simulated data using this test along with a thourogh investigation of its applicability can be found in \citet{Niels}.
	The study is organised as follows:
	Section \ref{method} presents our method and the general formalism we developed to evaluate the correlation functions analytically.
	Section \ref{sss:PP} details a complete example calculation for one of the correlation functions, namely $\langle P(\textbf{k}_{\perp}) \cdot \overline{P(\textbf{k}_{\perp}')} \rangle_{B}$.
	Section \ref{other} then  presents all the correlation functions up to fourth order in magnetic field strength.
	Section \ref{acidtest} introduces the \textit{LITMUS} test.
	Section \ref{con} presents finally a thorough evaluation of our findings.
	Details of the derivation of the other correlation functions are listed in the Appendix which also contains the remaining technical information regarding the study. 

\section{Methods}\label{method}
	We now proceed to calculate the correlation functions of $I(\fe{x}_{\perp})$, $P(\fe{x}_{\perp})$ and $\phi(\fe{x}_{\perp})$. Since all calculations resemble each other in respect of certain basic features, a general framework has been developed for them. Before presenting that, we introduce our basic notations and the magnetic correlation tensor, a quantity that will be referred to frequently but requires some preliminary explanation due to its complexity.
		\subsection{Notation}
			During this study we use the following definitions for the radio observables $I(\fe{x}_{\perp})$, $P(\fe{x}_{\perp})$ and $\phi(\fe{x}_{\perp})$:
			\begin{align}
			&I=\int dz \ \big(B_{1}^2+B_{2}^2\big), \\
			&P=\int dz \ \big(B_{1}^2-B_{2}^2+2iB_{1}B_{2}\big),\quad \text{and} \\
			&\phi=\int dz \ B_{3} .
			\end{align}
			Throughout this study, the coordinate axis $z$ always equals the line of sight.
			For convenience, all fore factors are suppressed including the electron density $n_e$, assumed to be constant. A detailed derivation can be found in Appendix \ref{sync}.

			Apart from this, we like to introduce some further notation we use frequently. In the following, the vectors $\fe{r}$ or $\fe{r}'$ shall always denote a combination such as $\fe{x}'-\fe{x}$ to be defined exactly when needed. Furthermore, $\fe{w}=(\fe{k}_{\perp}''',0), \fe{u}=(\fe{k}_{\perp}'',0)$, $\fe{v}=(\fe{k}_{\perp}',0)$ and $\textbf{a}=(-\textbf{q}_{\perp}-\textbf{k}_{\perp}',-q_{z})$ holds.

		\subsection{The magnetic correlation tensor}\label{M}
			We assume generally \textit{Gaussian statistics} for the magnetic field distribution in the observed region. This is, of course, a simplification but represents an initial starting point, especially as we are mainly interested in a proof of concept and therefore seek for simplicity. In addition, we assume \textit{statistical homogeneity} to first establish the most illuminating cases. This assumption is widely used in the literature. For an arbitrary field $\psi$, statistical homogeneity means that the two point correlation function of the field depends only on the distance of the two parts, $\langle\psi(\fe{x}') \psi^* (\fe{x})\rangle=C(\fe{r})$ with $\fe{r}=\fe{x}'-\fe{x}$. This automatically implies for this correlation function in Fourier space:
			\allowdisplaybreaks\begin{align*}
			\langle\psi(\fe{k}') \psi^* (\fe{k})\rangle&=(2\pi)^3 \delta^{3}(\fe{k}'-\fe{k}) P_{\psi} (k') \\
			\end{align*}
			where the $\psi$-power spectrum is specified by the Fourier transformed correlation function $P_{\psi} (k') \propto \int dr^3 C_{\psi}(r) \exp\left[ik'r\right]$  as stated by the Wiener-Kinchin-theorem \citep{2000fta..book.....B}.
			
			Within this study, the magnetic correlation tensor $M_{ij}(\fe{x},\fe{x}')=\langle B_{i}(\fe{x})B_{j}(\fe{x}')\rangle$ is frequently used for which this translational invariance leads to
			\begin{align}
			M_{ij}(\fe{x},\fe{x}')=M_{ij}(\fe{x}'-\fe{x}=\fe{r}) \quad \text{in normal space, and} 
			\end{align}
			\begin{align}
			\hat{M}_{ij}(\fe{k},\fe{k}')=(2\pi)^3 \delta^3(\fe{k}'-\fe{k}) \hat{M}_{ij}(\fe{k}') \quad \text{in Fourier space.} \label{transcon}
			\end{align}
			For homogeneous and isotropic magnetic turbulence the translationally invariant magnetic correlation tensor can be written as
			\begin{align}
			M_{ij}(\fe{r})=M_{N}(r)\delta_{ij}+\big(M_{L}(r)-M_{N}(r)\big)\frac{r_{i}r_{j}}{r^2}+M_{H}(r)\epsilon_{ijm}r_{m}\label{al:MO} 
			\end{align}
			with the longitudinal, normal and helical spectra denoted by $M_{L}(r)$, $M_{N}(r)$ and $M_{H}(r)$ respectively. The solenoidal condition $\nabla\cdot\fe{B}=0$ enables the connection of the two non-helical spectra by $M_{N}(r)=\frac{1}{2r}\frac{d}{dr}\big(r^2M_{L}(r)\big)$.
			By applying a Fourier transformation, we obtain:
			\begin{align}
			\hat{M}_{ij}(\fe{k})=\hat{M}_{N}(k)(\delta_{ij}-\frac{k_{i}k_{j}}{k^{2}})-i\epsilon_{ijm}
			\hat{H}(k) \frac{k_{m}}{k}\label{al:M} .
			\end{align}
			In this case, the condition $\nabla\cdot\fe{B}=0$ was used directly in the form $k_{i}\hat{M}_{ij}(\fe{k})=0$ to reduce the degrees of freedom to the normal and the helical spectra. These two functions are specified in terms of their real space counterparts as 
			\begin{align}
			&\hat{M}_{N}(k)=\int dr^3 M_{N}(r)\exp[i\fe{k}\fe{r}], \quad  \mbox{and} \\
			&\hat{H}(k)=\frac{d}{dk}\hat{M}_{H}(k)=\frac{d}{dk} \int dr^3 M_{H}(r)\exp[i\fe{k}\fe{r}].
			\end{align}
			Some interesting properties are:
			\begin{align}
			&M_{N}(0)=M_{L}(0) \quad \text{by definition,} \\
			&M_{ij}(0)=M_{N}(0) \ \delta_{ij}, \quad \text{and} \\
			&\hat{M}_{N}(0)=0, \quad \text{since} \ M_{N} \ \text{would diverge otherwise.}
			\end{align}
			The magnetic correlation tensor is closely related to the energy spectrum of the magnetic field. The field's mean energy density can be expressed as follows
			\begin{align}
			\frac{1}{8\pi} \langle \fe{B}^2(\fe{x}) \rangle &=\frac{1}{8\pi} \int \frac{dk^3}{(2\pi)^3} \Big\langle \sum_{i} B_{i}(\fe{k})B_{i}(\fe{k})\Big\rangle=\frac{1}{8\pi} \int \frac{dk^3}{(2\pi)^3}\sum_{i} \hat{M}_{ii}(\fe{k})= \notag\\
			&=\frac{1}{(2\pi)^3} \int_{0}^{\infty} dk \ k^2 \hat{M}_{N}(k)\overset{!}{=} \int_{0}^{\infty} dk \ \epsilon_{B}(k) \label{al:dens}.
			\end{align}
			In this case, we have used $M_{ii}(k)= 2 M_{N}$ and $\epsilon_{B}(k)$ denotes the 1D-energy density of $\fe{B}$.
			From this we derive 
			\begin{align}
			\epsilon_{B}(k) = \frac{k^{2} \hat{M}_{N}(k)}{8 \pi^{3}},
			\end{align}
			which is used in the following to replace $M_{N}(k)$ by the more commonly applied magnetic energy spectrum.
			
			Analogously to (\ref{al:dens}), we can relate the helical part of the spectrum $\hat{H}(k)$ to the mean \textit{current helicity} $\fe{j} \cdot \fe{B}$ and deduce a 1D-helical energy density $\epsilon_{H}(k)$:
			\begin{align}
			\langle \fe{j} \cdot \fe{B} \rangle&=\langle \fe{B} \cdot (\nabla \times \fe{B}) \rangle=\langle B_{l}(\fe{r}) \epsilon_{lij} \partial_{r_{i}} B_{j}(\fe{x}+\fe{r})\rangle|_{r=0}=\partial_{r_{i}} \epsilon_{lij} M_{lj}(r)|_{r=0} \notag\\
			&=i \epsilon_{lij} \int \frac{dk^3}{(2\pi)^3} k_{i} \hat{M}_{lj}(\fe{k})=\int \frac{dk^3}{(2\pi)^3} \epsilon_{lij} \epsilon_{ljm} \hat{H}(k) \frac{k_{i} k_{m}}{k} \notag\\
			&=-\int \frac{dk^3}{(2\pi)^3} \underbrace{\epsilon_{lji} \epsilon_{ljm}}_{2\delta_{im}} \hat{H}(k) \frac{k_{i} k_{m}}{k}=-\frac{8\pi}{(2\pi)^3} \int_{0}^{\infty} dk \ k^3 \hat{H}(k) \notag\\
			&\overset{!}{=} \int_{0}^{\infty} dk \ \epsilon_{H}(k). \label{curhel}
			\end{align}
			We can read off the 1D-helical energy density $\epsilon_{H}(k)$ which also can be used to substitute $H(k)$:
			\begin{align}
			\epsilon_{H}(k)=-\frac{k^3 \hat{H}(k)}{\pi^2}.\label{epsH}
			\end{align}
			
			Nevertheless, in our calculations we want to relate to the helicity spectrum $\hat{R}(k)$ rather then to the current helicity spectrum $\hat{H}(k)$ since it is the helicity $\fe{B} \cdot \fe{A}$ that usually is the subject of magnetohydrodynamics and dynamo theory.
			
			Fortunately, for isotropic, turbulent fields the current helicity and the magnetic helicity $\fe{B} \cdot \fe{A}$ are closely connected. Since the current helicity $(\nabla \times \fe{B}) \cdot \fe{B}$ has the same mathematical structure as the helicity $(\nabla \times \fe{A}) \cdot \fe{A}$, we can perform exactly the same derivation as in (\ref{curhel}) to show that the mean helicity relates to the helicity spectrum $\hat{R}(k)$ of the correlation tensor for the vector potential in the same way as the mean current helicity relates to the current helicity spectrum of the magnetic correlation tensor.
			We can construct the correlation tensor $\langle \fe{A}(\fe{x})\fe{A}^*(\fe{x}') \rangle$ for the magnetic vector potential $\fe{A}$ similar to the magnetic correlation tensor. If we assume translational invariance for the statistics and apply the Lorenz gauge condition $\nabla\cdot\fe{A}=0$ to the vector potential, all deductions made for the magnetic correlation tensor will also hold for the correlation tensor of the vector potential. In Fourier space it will have the form:
			\begin{equation}
			\langle \fe{A}_{m}(\fe{k})\fe{A}_{n}^*(\fe{k}') \rangle=\hat{R}_{mn}(\fe{k})=\hat{R}_{N}(k)(\delta_{mn}-\frac{k_{m}k_{n}}{k^{2}})-i\epsilon_{mnv} \hat{R}_{H}(k) \frac{k_{v}}{k}
			\end{equation}

			From here we can start by rewriting the magnetic correlation tensor in terms of the correlation tensor for the vector potential:
			\begin{equation}
			\langle \fe{B}_{i}(\fe{x}) \fe{B}_{j}^*(\fe{x}') \rangle = \epsilon_{ilm} \epsilon_{jrn} \partial_{\fe{x}_l} \partial_{\fe{x}_r} \langle \fe{A}_{m}(\fe{x}) \fe{A}_{n}^*(\fe{x}') \rangle
			\end{equation}
			If we now perfom a Fourier transformation on both sides and furthermore apply the condition for translational invariance (\ref{transcon}) we get:
			\begin{equation}
			\int \frac{dk^3}{(2\pi)^3} \hat{M}_{ij}(\fe{k}) = \int \frac{dk^3}{(2\pi)^3} \epsilon_{ilm} \epsilon_{jrn} k_{l} k_{r} \hat{R}_{mn}(\fe{k}) .
			\end{equation}
			Thus, we find
			\begin{equation}
			\hat{M}_{ij}(\fe{k}) = \epsilon_{ilm} \epsilon_{jrn} k_{l} k_{r} \hat{R}_{mn}(\fe{k}) \label{MR}.
			\end{equation}
			Using 
			\begin{equation}
			\epsilon_{ilm} \epsilon_{jrn} \epsilon_{mnv} = (\delta_{iv} \delta_{ln}-\delta_{in} \delta_{lv}) \epsilon_{jrn} = \delta_{iv}  \epsilon_{jrl}-\delta_{lv}  \epsilon_{jri}
			\end{equation}
			we can rewrite the right hand side of (\ref{MR}):
			\begin{align}
			\epsilon_{ilm} \epsilon_{jrn} k_{l} k_{r} \hat{R}_{mn}(\fe{k})=\underbrace{\epsilon_{ilm} \epsilon_{jrn} k_{l} k_{r} (\delta_{mn}-\frac{k_{m}k_{n}}{k^{2}})}_{\text{symmetric in i,j}}-\underbrace{k_{i} \epsilon_{jrl} k_{r} k_{l} \hat{R}_{H}(k)/k}_{=0}+\underbrace{i k^2 \epsilon_{ijr} k_{r} \hat{R}_{H}(k)/k}_{\text{antisymmetric in i,j}}
			\end{align}
			Finally, we find the relation between the current helicity spectrum $\hat{H}(k)$ and the helicity spectrum $\hat{R}_{H}$ by identifying the antisymmetric part of the left hand side of (\ref{MR}) with the antisymmetric part of the right hand side:
			\begin{align}
			-i\epsilon_{ijm} k_{m} \hat{H}(k)/k &= +i k^2 \epsilon_{ijr} k_{r} \hat{R}_{H}(k)/k \notag\\
			\Rightarrow \hat{H}(k) &= -k^2 \hat{R}_{H}(k) \label{finalcurhel}
			\end{align}
			From now on, although we mostly speak of helicity, we actually deal with current helicity for convenience and bear in mind that $\hat{H}(k)$ is easily convertible to $\hat{R}_{H}(k)$ using (\ref{finalcurhel}). \\
			A more detailed analysis of all these relations can be found in \citet{1978mfge.book.....M}.
			
			For the magnetic energy density in 1D Fourier space, a broken power-law is assumed in the following in our examples by adopting
			\begin{align}
			\epsilon_{B}(k)=\epsilon_{0}\Big(\frac{k}{k_{0}}\Big)^\beta \Big(1 +\Big(\frac{k}{k_{0}}\Big)^2\Big)^{-\frac{(\alpha+\beta)}{2}}\label{al:eps},
			\end{align}
			usually with $\beta=2$ and $k_{0}=1$ if not stated otherwise, but with different spectral indices $\alpha$. The low-$k$ asymptotic $\epsilon_{B}\approx k^2$ corresponds to a white noise spectrum without correlations on scales larger than $1/k_{0}$. For large $k$, we find $\epsilon_{B}\propto k^{-\alpha}$, eventually becoming a Kolmogorov-spectrum for $\alpha=5/3$.
			
			When ever necessary, we can always model the helicity power spectrum as $\hat{H}(k)=-\frac{\pi^2}{k^3} \epsilon_{H}(k)=\frac{\pi^2}{k^3} h(k) \epsilon_{B}(k)$, where $h(k)$ is a function between $-1$ and $1$. This can be seen from
			\begin{align}
			\hat{M}_{ij}(\fe{k})=\frac{\epsilon_{B}(k)}{k^{2}}\underbrace{\Big[8\pi(\delta_{ij}-\frac{k_{i}k_{j}}{k^{2}})-i\pi^2\epsilon_{ijm} h(k) \frac{k_{m}}{k}\Big]}_{A_{ij}} \label{al:MM}
			\end{align}
			and the fact that the matrix $A_{ij}$ must be positive definite. We adopt $\fe{k}=k e_{x}$ without loss of generality and find the characteristic polynomial of $A_{ij}$ to be
			\begin{align}
			&(1-\lambda)^2-h^2=0 \notag\\
			&\longrightarrow 1\pm h=\lambda \geq 0 \notag\\
			&\longrightarrow |h|\leq 1
			\end{align}
			This yields that $h \in \left[-1,1 \right]$.

		\subsection{General framework for all calculations} \label{sec:gen}
			The correlation functions of our observables can be calculated in a general and consistent way which we want to present now.
			
			Before we start, some general remarks about the mathematics are in place.
			Throughout this study an expression such as $\textbf{J}^{\dagger}\textbf{B}$ relates to a multidimensional scalar product:
			\begin{align}
			\textbf{J}^{\dagger}\textbf{B}=\sum_i \int dx^3 \ \fe{J}_{i}^* (x) \fe{B}_{i}(x).\label{multiscalar}
			\end{align}
			This definition includes a discrete summation over indices as well as a continuous integral over position space. The symmetric properties of matrix objects defined over a space with a scalar product (\ref{multiscalar}) reflect the appearance of discrete summation and continous integration. Therefore, a matrix element $M_{ij}(\fe{x},\fe{y})$ is called symmetric (or hermitian for complex quantities), if it is symmetric under a transposition of its indices \textit{and} under an interchange of its vectors $\fe{r}$:
			\begin{equation}
			M_{ij}^{\dagger}(\fe{x},\fe{y})=M_{ji}(\fe{y},\fe{x}) \label{al:hersym}.
			\end{equation}
			Thus, a symmetrised element is expressed as
			\begin{equation}
			M_{ij,\text{sym}}(\fe{r})=\frac{1}{2}\Big(M_{ij}(\fe{r})+M_{ji}(-\fe{r})\Big)\label{al:sym},
			\end{equation}
			where $\fe{r}=\fe{y}-\fe{x}$. In the case where a matrix element is only symmetrised for index transposition, we call it \textit{index-symmetric}:
			\begin{equation}
			M_{ij,\mathrm{isym}}(\fe{x},\fe{y})=\frac{1}{2}\Big(M_{ij}(\fe{r})+M_{ji}(\fe{r})\Big)\label{al:insym}.
			\end{equation}
			This distinction between symmetric and index-symmetric is important, because it takes care of subtleties that could easily generate confusion. We like to emphasize the difference between both symmetry operations, when applied to the magnetic correlation tensor (\ref{al:M}). The tensor contains an intrinsic symmetric and an intrinsic antisymmetric element. Regarding (\ref{al:sym}), the intrinsic antisymmetric part is preserved, whereas regarding (\ref{al:insym}) it is not. This is of paramount relevance, since information on the helical power spectrum is only preserved, if the intrinsic antisymmetric parts do not cancel out during calculations.
			
			Furthermore, we like to introduce the functional derivative, which is the natural generalisation of a derivative to function vector spaces. Its precise definition is \citep[see][]{1995iqft.book.....P}:
			\begin{align}
			&\frac{\delta}{\delta \textbf{J}_{i}(\textbf{x})}\textbf{J}_{j}(\textbf{y})=\delta^3(\textbf{x}-\textbf{y}) \delta_{ij}& \notag\\
			&\frac{\delta}{\delta \textbf{J}_{i}(\textbf{x})}\half \textbf{J}^{\dagger}\textbf{M}\textbf{J}=\frac{\delta}{\delta \textbf{J}_{i}(\textbf{x})} \int dy^{3} \int dy'^{3} \frac{1}{2}\textbf{J}_{k}(\textbf{y})\textbf{M}_{kl}(\textbf{y},\textbf{y}')\textbf{J}_{l}(\textbf{y}')& \notag\\
			&=\int dy'^{3}\bigg[\frac{1}{2}\fe{M}_{il}(\fe{x},\fe{y}')\fe{J}_{l}(\fe{y}')+\frac{1}{2}\fe{J}_{k}(\fe{y}')\fe{M}_{ki}(\fe{x},\fe{y}')\bigg]& \notag\\ 
			&=\frac{1}{2} (\textbf{J}^{\dagger}\textbf{M})_{i}(\textbf{x})+\frac{1}{2}(\fe{M}\fe{J})_{i}(\fe{x})& \notag\label{funcdev}
			\end{align}
			For convenience and to avoid confusion with the delta function, we sometimes adopt easier notations:
			\begin{align}
			\frac{\delta}{\delta \textbf{J}_{i}(\textbf{x})}=\partial_{J_{i}}(\textbf{x})=\partial_{i}(\textbf{x})
			\end{align}
			
			Now we proceed, presenting the framework of the calculations. The general evaluation of the expectation value of a function $X$ of observables for Gaussian magnetic field statistics with covariance matrix $M$ and its determinant $|M|$ is conducted as follows\footnote{We denote with $X$ either $I$, $P$ or $\phi$ or combinations thereof up to fourth order in the magnetic field in Fourier space. Therefore $X$ contains up to 4 Fourier vectors within the observed plane labeled with primes. We denote with $F$ the real space source function of $X$ which depends directly on the local components of the magnetic field $B_{i}$ so that $X=\int dz \ldots \int dz''' F(\fe{B},\fe{B}',\fe{B}'',\fe{B}''')$.}:
			
			\begin{align}
			& \langle X(\textbf{k}_{\perp},\textbf{k}_{\perp}',...) \rangle_{\textbf{B}}=\vfactor \int \mathcal{D}B \ X(\textbf{k}_{\perp},\textbf{k}_{\perp}',...) \exp[\magnet] \notag\\
			&=\vfactor \int \mathcal{D}B \int d\textbf{x}_{\perp} \ ...\int d\textbf{x}_{\perp}''' \int dz \ ...\int dz''' \ F(B_{i}(\textbf{x}),B_{j}(\textbf{x}'),...) \notag\\
			&\quad \  \exp[\magnet] \ \exp[i(\textbf{k}_{\perp}\textbf{x}_{\perp}+\textbf{k}_{\perp}'\textbf{x}_{\perp}'+...)] \notag\\
			&=\vfactor \int \mathcal{D}B \int d\textbf{x}_{\perp} \ ...\int d\textbf{x}_{\perp}''' \int dz \ ...\int dz''' \ F\left(\partial_{J_{i}}(\textbf{x}),\partial_{J_{j}}(\textbf{x}'...)\right) |_{\textbf{J}=0} \notag\\
			&\quad \ \exp[\magnet+\textbf{J}^{\dagger}\textbf{B}] \ \exp[i(\textbf{k}_{\perp}\textbf{x}_{\perp}+\textbf{k}_{\perp}'\textbf{x}_{\perp}'+...)]\label{al:squ} \\
			&=\int d\textbf{x}_{\perp} \ ...\int d\textbf{x}_{\perp}''' \int dz \ ...\int dz''' \ \exp[i(\textbf{k}_{\perp}\textbf{x}_{\perp}+\textbf{k}_{\perp}'\textbf{x}_{\perp}'+...)] \notag\\ 
			&\quad \ F\left(\partial_{J_{i}}(\textbf{x}),\partial_{J_{j}}(\textbf{x}'...)\right) |_{\textbf{J}=0} \ \exp[\frac{1}{2}\textbf{J}^{\dagger}\textbf{M}\textbf{J}] \notag\\
			&=\int d\textbf{x} \ ...\int d\textbf{x}''' \ \exp[i(\textbf{k}_{\perp}\textbf{x}_{\perp}+\textbf{k}_{\perp}'\textbf{x}_{\perp}'+...)] \notag\\ &\quad \ F\left(\partial_{J_{i}}(\textbf{x}),\partial_{J_{j}}(\textbf{x}'...)\right) |_{\textbf{J}=0} \ \exp[\frac{1}{2}\textbf{J}^{\dagger}\textbf{M}\textbf{J}] \label{al:gen}
			\end{align}
			At this point, it is necessary to consider the actual form of $F$. The calculations vary enormously as complexity of the correlation functions differs. From now on we identify the covariance matrix of the Gaussian distribution with the magnetic correlation tensor. This identity is proven in Appendix \ref{cocor}. In (\ref{al:squ}) we introduced a generating functional $\fe{J}$ and completed the square of the exponential. By integrating out the shifted Gaussian part, which depends on $\fe{B}$, we are left with the part solely dependent on $\fe{J}$:
			\begin{flalign}
			& \vfactor \int \mathcal{D}B \exp[\magnet+\textbf{J}^{\dagger}\textbf{B}]& \notag\\
			&=\vfactor \int \mathcal{D}B \exp[\magnet+\frac{1}{2} \textbf{J}^{\dagger}\textbf{B}+\frac{1}{2}\textbf{B}^{\dagger}\textbf{J}]& \notag\\
			&=\vfactor \int \mathcal{D}B \exp[\magnet+\frac{1}{2} \textbf{J}^{\dagger}\textbf{M}\textbf{M}^{-1}\textbf{B}+\frac{1}{2}\textbf{B}^{\dagger}\textbf{M}^{-1}\textbf{M}\textbf{J}]& \notag\\
			&=\vfactor \int \mathcal{D}B \exp[-\frac{1}{2}(\textbf{B}-\textbf{M}\textbf{J})^{\dagger}\textbf{M}^{-1}(\textbf{B}-\textbf{M}\textbf{J})] \ \exp[\frac{1}{2} \textbf{J}^{\dagger}\textbf{M}\textbf{J}]& \notag\\
			&=\exp[\frac{1}{2} \textbf{J}^{\dagger}\textbf{M}\textbf{J}]&
			\end{flalign}
			In (\ref{al:gen}), we finally made the assumption that the observed space is sufficiently extended, so that we can neglect the finiteness of the integrals over the lines of sight $\int dz$ and treat them as if they were infinite. 
			
			Introducing the generating functional $\fe{J}$ and thereby changing the fields $\fe{B}(\fe{x})$ to the functional derivative  $\partial_{\fe{J}_{i}}(\fe{x})=\frac{\partial}{\partial\fe{J}_{i}}(\fe{x})$
			in (\ref{al:squ}) provides a powerful method to calculate the integral $\mathcal{D}B$ over all possible magnetic field configurations as an infinite-dimensional path integral. With the definition of the functional derivative at hand (\ref{funcdev}), we can discuss its actual evaluation.
			Since all uneven products of functional derivatives in (\ref{al:gen}) give zero because they also leave expressions with $\fe{J}$  which have been ``taken down'' from the exponential function during the differentiation, we are left with just two general types of possible combinations: Two or four derivatives.
			
			In addition to the generating functional technique familiar from quantum field theory, we also apply the renowned Wick theorem \citep[see for example][]{1995iqft.book.....P} to evaluate the remaining derivatives in an elegant, quick and safe manner, rather than calculating them by brute force. The Wick theorem can be used under conditions which will become clear if one looks at how the underlying differentiation works. Firstly, we need the covariance matrix to be symmetric or hermitian. This, as already mentioned, means 
			\begin{align}
			M_{ij}(\fe{x},\fe{y})=M_{ij}^{\dagger}(\fe{x},\fe{y})=M_{ji}(\fe{y},\fe{x})
			\end{align}
			which is fulfilled by (\ref{al:MO}) and (\ref{al:M}). However, since there will be one case where the covariance matrix is not symmetric or hermitian (see Chapter \ref{sss:PP}) and since we have to take thorough care of the exact order of the vectors, discussed in more detail below, the Wick theorem is expressed in a form that takes care of these subtleties:
			\begin{flalign} 
			&\partial_{i}(\fe{a}) \partial_{j}(\fe{b})\left.\exp\source\right|_{\fe{J}=0}& \notag\\
			&=\partial_{i}(\fe{a})\left[\jm{j}{b}+\mj{j}{b}\right] \left.\exp\source\right|_{\fe{J}=0}& \notag\\
			&=\half \fe{M}_{ij}(\fe{a},\fe{b})+\half \fe{M}_{ji}(\fe{b},\fe{a})& \\
			\intertext{and}\notag
			&\partial_{i}(\fe{a})\partial_{j}(\fe{b})\partial_{k}(\fe{c})\partial_{l}(\fe{d})\left.\exp\source\right|_{\fe{J}=0}& \notag\\
			&=\partial_{i}(\fe{a})\partial_{j}(\fe{b})\partial_{k}(\fe{c})\left[\jm{l}{d}+\mj{l}{d}\right]\left.\exp\source\right|_{\fe{J}=0}& \notag\\
			&=\partial_{i}(\fe{a})\partial_{j}(\fe{b})\bigg[\Big(\half \fe{M}_{kl}(\fe{c},\fe{d})+\half \fe{M}_{lk}(\fe{d},\fe{c})\Big)& \notag\\
			&+\Big(\mj{l}{d}+\jm{l}{d}\Big)\Big(\jm{k}{c}+\mj{k}{c}\Big)\bigg] \left.\exp\source\right|_{\fe{J}=0}& \notag\\
			&=\partial_{i}(\fe{a})\bigg[\Big(\half \fe{M}_{kl}(\fe{c},\fe{d})+\half \fe{M}_{lk}(\fe{d},\fe{c})\Big)\Big(\jm{j}{b}+\mj{j}{b}\Big)& \notag\\
			&+\Big(\half \fe{M}_{lj}(\fe{d},\fe{b})+\half \fe{M}_{jl}(\fe{b},\fe{d})\Big)\Big(\jm{j}{c}+\mj{j}{c}\Big)& \notag\\
			&+\Big(\jm{l}{d}+\mj{l}{d}\Big)\Big(\half \fe{M}_{kj}(\fe{c},\fe{b})+\half \fe{M}_{jk}(\fe{b},\fe{c}\Big)+\ldots\bigg] \left.\exp\source\right|_{\fe{J}=0}& \notag\\
			&=\Big(\mm{kl}{c}{d}{lk}\Big)\Big(\mm{ij}{a}{b}{ji}\Big) &\notag\\
			&+\Big(\mm{lj}{d}{b}{jl}\Big)\Big(\mm{ki}{c}{a}{ik}\Big) &\notag\\
			&+\Big(\mm{il}{a}{d}{li}\Big)\Big(\mm{kj}{c}{b}{jk}\Big). &\label{al:dif}
			\end{flalign}
			Now all derivatives up to fourth order can be calculated just by simply inserting the current case. For example $\partial_{1}^{2}(\fe{x})\partial_{1}^{2}(\fe{x}')|_{\fe{J}=0}\exp\source$ can be read off from (\ref{al:dif}) by inserting  $i=j=k=l=1, \fe{a}=\fe{b}=\fe{x}\ \text{and} \ \fe{c}=\fe{d}=\fe{x}'$ :
			\begin{align}
			&\partial_{1}^{2}(\fe{x})\partial_{1}^{2}(\fe{x}')\left.\exp\source\right|_{\fe{J}=0}=M_{11}^2(0)+2M_{11}^{2}(\fe{r})
			\end{align}
			As we can see from (\ref{al:dif}), all antisymmetric parts cancel out during the differentiation. Thus although $\fe{M}$ was not explicitly restricted to be symmetric, only the symmetric elements of the magnetic correlation tensor
			\begin{align}
			M_{ij,\text{sym}}(\fe{r})=\frac{1}{2}\Big(M_{ij}(\fe{r})+M_{ji}(-\fe{r})\Big)
			\end{align}
			remain in the end. However, it is important to understand that these symmetric elements actually preserve the intrinsic antisymmetric parts that constitute the magnetic correlation tensor. As mentioned above, this is because we take into account the inversion of the vector $\fe{r}$ when transposing the tensor elements. A look at (\ref{al:MO}) reveals, that the minus sign of the Levi-Civita-tensor $\epsilon_{ijm}$ we encounter under interchanged indices is exactly cancelled by the minus sign occuring due to inversion of the vector $\fe{r}$:
			\begin{align}
			M_{ij}(\fe{r})=M_{N}(r)\delta_{ij}+\big(M_{L}(r)-M_{N}(r)\big)\frac{r_{i}r_{j}}{r^2}+M_{H}(r)\epsilon_{ijm}r_{m}. \label{again}
			\end{align}
			This means that although the tensor (\ref{again}) is symmetric in the general way defined in (\ref{al:sym}), it is not index-symmetric due to its individual antisymmetric constituents. The consequence is that if we carry out the derivatives using the Wick theorem we not only have to take care of the right combination of indices but also of the corresponding vectors $\fe{r}$ or $-\fe{r}$ and, in the end only the symmetric parts of $\fe{M}$, as defined in (\ref{al:sym}), appear. And further, this does mean that if we encounter index-symmetric expressions such as $M_{ij}(\fe{r})=\big(M_{ij}(\fe{r})+M_{ji}(\fe{r})\big)/2$, the intrinsic antisymmetric part related to the helical power spectrum in (\ref{again}) is lost during the differentiation. A careful look at (\ref{al:dif}) reveals that with the right combination for $i,j,k,l$ and $\fe{a},\fe{b},\fe{c}, \fe{d}$ and a sum of terms as in (\ref{al:dif}), it is possible to get such combinations. For example
			\begin{align}
			\bigg(&\partial_{1}^2(\fe{x})\partial_{2}^2(\fe{x}')+\partial_{2}^2(\fe{x})\partial_{1}^2(\fe{x}')\bigg) \exp\source|_{\fe{J}=0}=2M_{11}(0)M_{22}(0)\notag\\
			&+2\big(M_{21}^2(\fe{x-x'})+M_{12}^2(\fe{x-x'})\big)=2M_{11}(0)M_{22}(0)+4M_{21,\mathrm{isym}}^2(\fe{x-x'})\label{al:symcobo}
			\end{align}

\section[Calculational example]{The polarisation 2-point function $\langle P(\textbf{k}_{\perp}) \cdot P^*(\textbf{k}_{\perp}') \rangle_{B}$}\label{sss:PP}
	In this section the polarisation 2-point function $\langle P(\textbf{k}_{\perp}) \cdot P^*(\textbf{k}_{\perp}') \rangle_{B}$ is calculated to serve us as an example for the general calculation to obtain the other correlation functions of our observables. Since the steps are similar for all correlation functions and differ only in complexity, we intend to present them in detail only for a single case here and just list the other calculations in the Appendix \ref{ap:cal}.
	
\begin{figure*} [!h]
	\centering
	\includegraphics[width=0.46\textwidth]{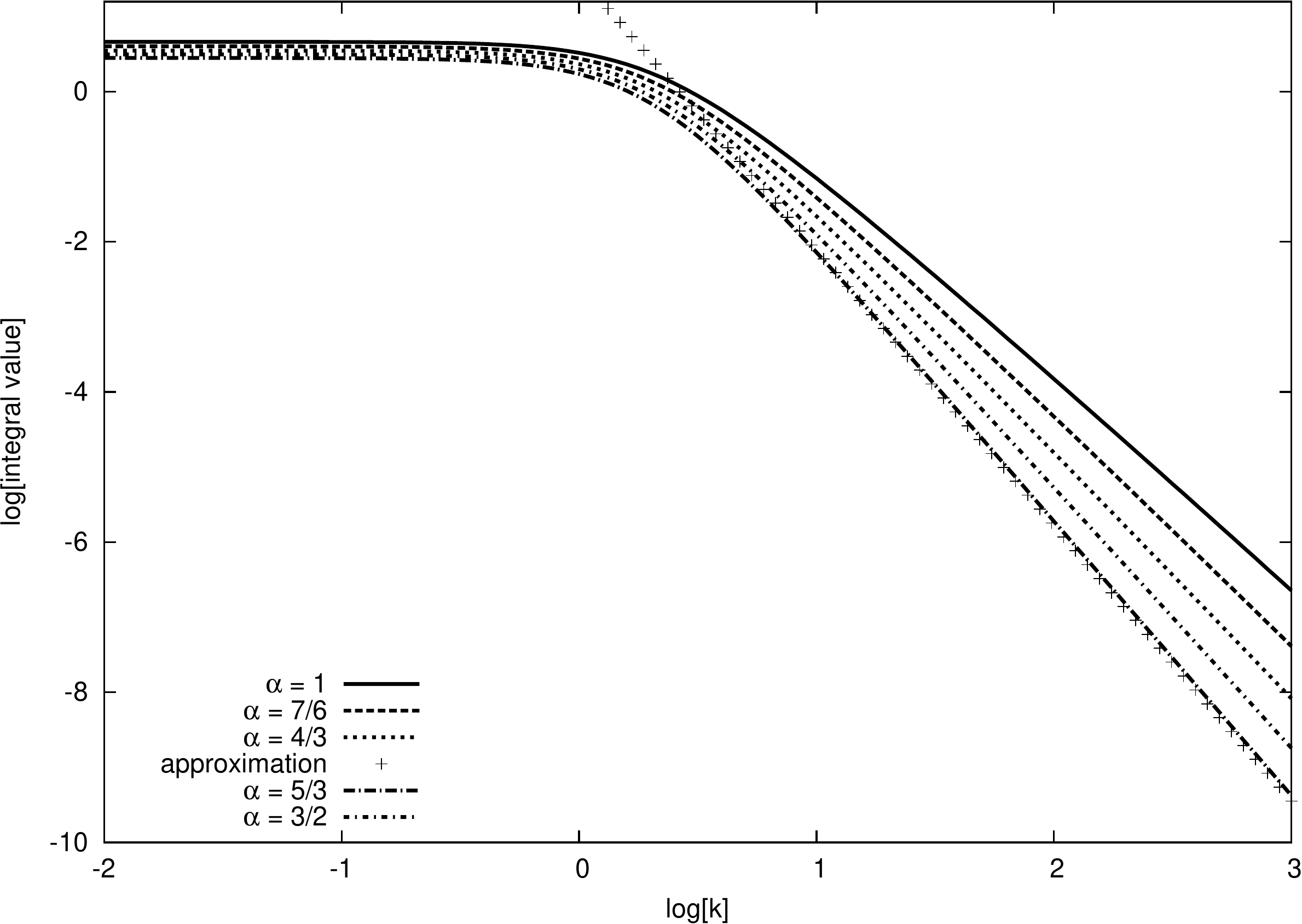}\includegraphics[width=0.46\textwidth]{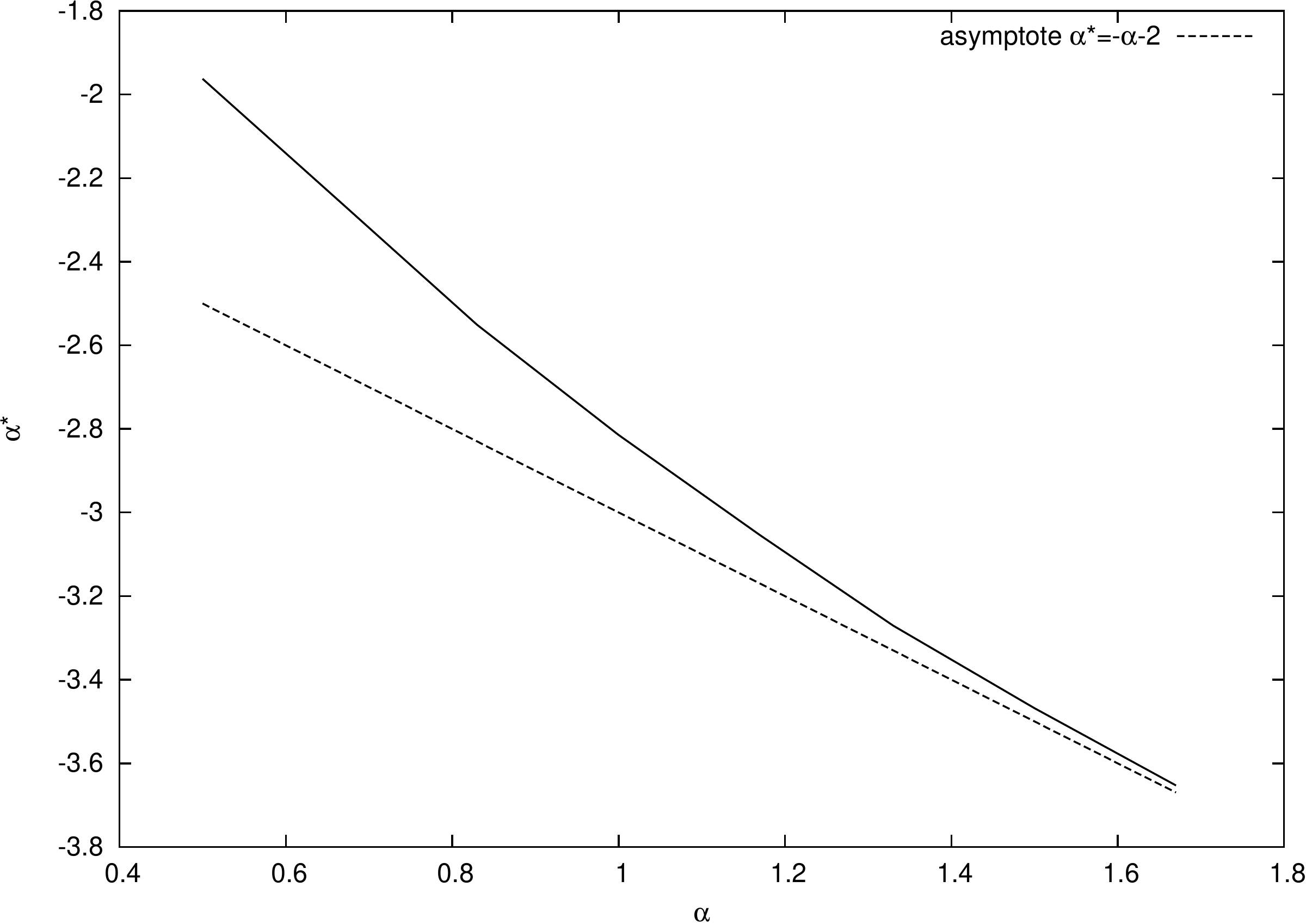}
	\caption{\textbf{Left:} Numerically evaluated integral values of $\langle P(\textbf{k}_{\perp}) \cdot \overline{P(\textbf{k}_{\perp}')} \rangle_{B}$ in a log-log diagram with an approximation for the $\alpha=5/3$ case. \textbf{Right:} Plot of the energy spectrum slopes $\alpha$ against the polarisation spectrum slopes $\alpha*$ in $\langle P(\textbf{k}_{\perp}) \cdot \overline{P(\textbf{k}_{\perp}')} \rangle_{B}$.}
	\label{bild2}
	\label{bild1}
	\end{figure*}

	As $\langle P(\textbf{k}_{\perp}) \cdot P^* (\textbf{k}_{\perp}') \rangle_{B}$ is of fourth order in the magnetic field and, in addition, $P$ has a rather complex dependence on $\bf{B}$, it is convenient to introduce a compact notation for $P(\textbf{x}_{\perp})=\int_{0}^{L} dz (B_{1}(\textbf{x})+iB_{2}(\textbf{x}))^{2}$ in order to clarify the calculation as far as possible. Defining $B_{\pm}=\frac{1}{\sqrt{2}}(B_{1} \pm iB_{2})$ allows the expression $P(\fe{x}_{\perp})=\int_{0}^{L} dz \ 2B_{+}^{2}(\fe{x})$ and $P^*(\fe{x}_{\perp})=\int_{0}^{L} dz \ 2B_{-}^{2}(\fe{x})$. Thus, a change of basis of $\textbf{B}$ is introduced, mapping $\textbf{B}=(B_{1},B_{2},B_{3}) \longrightarrow \tilde{\textbf{B}}=(B_{+},B_{-},B_{3})$. We can then work effectively with $2B_{+}^{2}(\textbf{x}) \cdot 2B_{-}^{2}(\textbf{x}')$ instead of $(B_{1}(\textbf{x})+iB_{2}(\textbf{x}))^{2} \cdot (B_{1}(\textbf{x}')-iB_{2}(\textbf{x}'))^{2}$. With regard to the correlation function, this results in:
	\begin{align}
	\langle P(\textbf{k}_{\perp}) \cdot P^*(\textbf{k}_{\perp}') \rangle_{B} =&\ 4 \int d\textbf{x}_{\perp} \int d\textbf{x}_{\perp}' \int dz \int dz'
	\exp[\fourierI] \notag\\
	& \partial_{J_{+}}^{2}(\textbf{x}) \partial_{J_{-}}^{2}(\textbf{x}') \left.\exp\left[\frac{1}{2}\textbf{J}^{\dagger}\textbf{M}\textbf{J}\right]\right|_{\textbf{J}=0} \label{al:pol}
	\end{align}
	Regarding the differentiation with respect to $J_{\pm}$ we need to establish a relation between $J_{1/2}$ and $J_{\pm}$. The basis transformation should preserve all scalar products, therefore, we have $J_{+}^{\dagger}B_{+}+J_{-}^{\dagger}B_{-}+J_{3}^{\dagger}B_{3}=J_{1}^{\dagger}B_{1}+J_{2}^{\dagger}B_{2}+J_{3}^{\dagger}B_{3}$. Using this we establish the required relations:
	\begin{align}
	& J_{+}^{+}B_{+}+J_{-}^{\dagger}B_{-} = J_{1}^{\dagger}B_{1}+J_{2}^{\dagger}B_{2}  \\
	& \longrightarrow J_{1}^{\dagger}=\frac{1}{\sqrt{2}}(J_{+}^{\dagger}+J_{-}^{\dagger}) \ \text{and} \ J_{2}^{\dagger}=\frac{i}{\sqrt{2}}(J_{+}^{\dagger}-J_{-}^{\dagger}) \notag\\
	& \text{from which we obtain} \quad J_{\pm}=\frac{1}{\sqrt{2}}(J_{1} \pm iJ_{2}) \\
	& \text{and also} \quad J_{1}=\frac{1}{\sqrt{2}}(J_{-}+J_{+}) \quad \text{as well as} \quad J_{2}=\frac{i}{\sqrt{2}}(J_{-}-J_{+}).
	\end{align}
	Thus, the transformation matrices $\textbf{J}=\textbf{O}\tilde{\textbf{J}}$ and $\tilde{\textbf{J}}=\textbf{O}^{\dagger}\textbf{J}$ are:
	$$O = \begin{pmatrix} \frac{1}{\sqrt{2}} & \frac{1}{\sqrt{2}} & 0 \\ -\frac{i}{\sqrt{2}} & \frac{i}{\sqrt{2}} & 0 \\ 0 & 0 & 1 \end{pmatrix}, \quad \text{and} \ O^{\dagger} = \begin{pmatrix} \frac{1}{\sqrt{2}} & \frac{i}{\sqrt{2}} & 0 \\ \frac{1}{\sqrt{2}} & -\frac{i}{\sqrt{2}} & 0 \\ 0 & 0 & 1 \end{pmatrix}$$
	We now need to express the argument $\textbf{J}^{\dagger}\textbf{M}\textbf{J}$ of the exponential in (\ref{al:pol}) in terms of the transformed quantities:
	$$\textbf{J}^{\dagger}\textbf{M}\textbf{J}=\tilde{\textbf{J}}^{\dagger}\textbf{O}^{\dagger}\textbf{M}\textbf{O}\tilde{\textbf{J}}=\tilde{\textbf{J}}^{\dagger}\tilde{\textbf{M}}\tilde{\textbf{J}} \quad \quad \text{with} \quad \tilde{\textbf{M}}=\textbf{O}^{\dagger}\textbf{M}\textbf{O}$$
	Some elements of $\tilde{\textbf{M}}$ that will soon become important are:
	\begin{align}
	&\tilde{M}_{++}(r)=\frac{1}{2}\left(M_{11}(r)+M_{22}(r)-iM_{12}(r)+iM_{21}(r)\right) \\
	&\tilde{M}_{--}(r)=\frac{1}{2}\left(M_{11}(r)+M_{22}(r)+iM_{12}(r)-iM_{21}(r)\right) \\
	&\tilde{M}_{+-}(r)=\frac{1}{2}\left(M_{11}(r)-M_{22}(r)+iM_{12}(r)+iM_{21}(r)\right) \\ &\tilde{M}_{-+}(r)=\frac{1}{2}\left(M_{11}(r)-M_{22}(r)-iM_{12}(r)-iM_{21}(r)\right)
	\end{align}
	The entire matrix then reads:
	$$ \tilde{\textbf{M}}= \begin{pmatrix} \tilde{M}_{++}(r) & \tilde{M}_{+-}(r) & \frac{1}{\sqrt{2}}(M_{13}+M_{23}) \\ 
	\tilde{M}_{-+}(r) & \tilde{M}_{--}(r) & \frac{1}{\sqrt{2}}(M_{13}-M_{23}) \\ \frac{1}{\sqrt{2}}(M_{31}-iM_{32}) &  \frac{1}{\sqrt{2}}(M_{31}+iM_{32}) & M_{33} \end{pmatrix}
	$$
	Returning to the correlation function (\ref{al:pol}), we can now carry out the functional derivatives. Since $\tilde{\fe{J}}$ is a complex quantity, we now have to concern ourselves with the complex conjugation implied in the $\dagger$ operation which affects $\tilde{\fe{J}}^*=(J_{+}^*,J_{-}^*,J_{3}^*)=(J_{-},J_{+},J_{3})$. Using (\ref{al:dif}) we find  :
	\begin{align}
	& \partial_{J_{+}}(\textbf{x}) \ \partial_{J_{+}}(\textbf{x}) \ \partial_{J_{-}}(\textbf{x}') \ \partial_{J_{-}}(\textbf{x}') \ \exp[\frac{1}{2}\Ji^{\dagger} \tilde{\textbf{M}} \Ji] \ |_{\Ji=0}= 2\big(M_{11}(\textbf{r})+M_{22}(\textbf{r})\big)^{2}
	\end{align}
	Inserting this into the overall equation for the correlation function (\ref{al:pol}) results in:
	\begin{align}
	&\langle P(\textbf{k}_{\perp}) \cdot P^*(\textbf{k}_{\perp}')\rangle
	= 8 \int dx^{3} \int dx^{3'} \big(M_{11}(\textbf{r})+M_{22}(\textbf{r})\big)^2 \exp[\fourierI] \notag\\ 
	& = 8 \int dx^{3} \int dr^{3}  \cdot \big(M_{11}^{2}(\textbf{r})+M_{22}^{2}(\textbf{r})+2M_{22}(\textbf{r})M_{11}(\textbf{r})\big)\exp[i\textbf{x}_{\perp}(\textbf{k}_{\perp}-\textbf{k}_{\perp}')] \notag\\
	&\quad \ \exp[-i\textbf{r}_{\perp}\textbf{k}_{\perp}'] \notag\\
	& = 8 \int dx^{3} \int dr^{3} \int \frac{dq^{3}}{(2\pi)^{3}} \int \frac{dq'^{3}}{(2\pi)^{3}} \cdot \big(\hat{M}_{11}(\textbf{q})\hat{M}_{11}(\textbf{q}')+\hat{M}_{22}(\textbf{q})\hat{M}_{22}(\textbf{q}') \notag\\
	&\quad \ +\hat{M}_{22}(\textbf{q})\hat{M}_{11}(\textbf{q}')+\hat{M}_{22}(\textbf{q}')\hat{M}_{11}(\textbf{q})\big) \exp[i\textbf{x}_{\perp}(\textbf{k}_{\perp}-\textbf{k}_{\perp}')] \exp[-i\textbf{r}(\textbf{q}+\textbf{q}')] \notag\\
	&\quad \ \exp[-i\textbf{r}_{\perp}\textbf{k}_{\perp}'] \notag\\
	& = 8 (2\pi)^{2} \delta^{2}(\textbf{k}_{\perp}-\textbf{k}_{\perp}') \int dz \int dr^{3} \int \frac{dq^{3}}{(2\pi)^{3}} \int \frac{dq'^{3}}{(2\pi)^{3}} \cdot \big(\hat{M}_{11}(\textbf{q})\hat{M}_{11}(\textbf{q}') \notag\\
	&\quad\ +\hat{M}_{22}(\textbf{q})\hat{M}_{22}(\textbf{q}')+\hat{M}_{22}(\textbf{q})\hat{M}_{11}(\textbf{q}')+\hat{M}_{22}(\textbf{q}')\hat{M}_{11}(\textbf{q})\big) \exp[-i\textbf{r}_{\perp}(\textbf{q}_{\perp}+\textbf{q}_{\perp}'+\textbf{k}_{\perp}')] \notag\\
	&\quad \ \exp[-ir_{z}(q_{z}+q_{z}')] \notag\\
	& = \frac{8}{2\pi} \delta^{2}(\textbf{k}_{\perp}-\textbf{k}_{\perp}') L_{z} \int dq^{3} \int dq'^{3} \cdot \delta^{2}(\textbf{q}_{\perp}+\textbf{q}_{\perp}'+\textbf{k}_{\perp}') \delta(q_{z}+q_{z}') \big(\hat{M}_{11}(\textbf{q})\hat{M}_{11}(\textbf{q}') \notag\\
	&\quad\ +\hat{M}_{22}(\textbf{q})\hat{M}_{22}(\textbf{q}')+\hat{M}_{22}(\textbf{q})\hat{M}_{11}(\textbf{q}')+\hat{M}_{22}(\textbf{q}')\hat{M}_{11}(\textbf{q})\big) \notag\\
	&\overset{\textbf{a}=(-\textbf{q}_{\perp}-\textbf{k}_{\perp}',-q_{z})} {=} \frac{8}{2\pi} \delta^{2}(\textbf{k}_{\perp}-\textbf{k}_{\perp}') L_{z} \int dq^{3} \cdot \big(\hat{M}_{11}(\textbf{q})\hat{M}_{11}(\textbf{a})+\hat{M}_{22}(\textbf{q})\hat{M}_{22}(\textbf{a}) \notag\\
	&\quad \ +\hat{M}_{22}(\textbf{q})\hat{M}_{11}(\textbf{a})+\hat{M}_{22}(\textbf{a})\hat{M}_{11}(\textbf{q})\big) \notag\\
	& = \frac{8}{2\pi} (2 \pi)^{6} \delta^{2}(\textbf{k}_{\perp}-\textbf{k}_{\perp}') L_{z} \int dq^{3} \cdot \frac{\epsilon_{B}(q) \epsilon_{B}(a)}{q^{2}a^{2}} \cdot \Big[\Big(1-\frac{q_{x}^2}{q^{2}}\Big)\Big(1-\frac{a_{x}^2}{a^{2}}\Big) \notag\\
	&\quad \ +\Big(1-\frac{q_{y}^2}{q^{2}}\Big)\Big(1-\frac{a_{y}^2}{a^{2}}\Big)+\Big(1-\frac{q_{y}^2}{q^{2}}\Big)\Big(1-\frac{a_{x}^2}{a^{2}}\Big)
	+\Big(1-\frac{q_{x}^2}{q^{2}}\Big)\Big(1-\frac{a_{y}^2}{a^{2}}\Big)\Big] \notag\\
	&=8(2 \pi)^{5} \delta^{2}(\textbf{k}_{\perp}-\textbf{k}_{\perp}') L_{z} \underbrace{\int dq^{3} \cdot \frac{\epsilon_{B}(q) \epsilon_{B}(a)} {q^{2}a^{2}} \Big[\Big(2-\frac{q_{\perp}^{2}}{q^{2}}\Big)\Big(2-\frac{a_{\perp}^{2}}{a^{2}}\Big)\Big]}_{K} \label{alig:int}
	\end{align}
	This integral can be further simplified. To do this, we transform it into spherical coordinates and perform the subintegral over $\varphi$ analytically. The remaining 2-dimensional integral can be done numerically without problems. We choose the axes to be selected so that the angle $\theta$ is between the $x$-axis and vector $\textbf{q}_{\perp}$ while the angle $\varphi$ rotates around the $x$-axis. Without loss of generality, we choose  $\textbf{k}_{\perp}=k_{\perp} e_{x}$ to ensure that the angle between $\textbf{q}_{\perp}$ and $\textbf{k}_{\perp}$ coincides with $\theta$.
	Thus the transformation is 
	\begin{align}
	&q_{x}=q \cos{\theta} \\
	&q_{y}=q \sin{\theta}\sin{\varphi} \\
	&q_{z}=q \sin{\theta}\cos{\varphi}.
	\end{align}
	Which implies for the required quantities:
	\begin{align}
	q_{\perp}^2&=q_{x}^2+q_{y}^2=q^2\sin^2\theta\sin^2\varphi+\cos^2\theta=q^2(1-\sin^2\theta\cos^2\varphi) \\
	a_{\perp}^2&=(\textbf{q}_{\perp}+\textbf{k}_{\perp})^2=q_{\perp}^2+k_{\perp}^2+2 \ \textbf{q}_{\perp} \cdot \textbf{k}_{\perp}=q_{\perp}^2+k_{\perp}^2+2qk_{\perp}\cos\theta \\
	a^2&=(\textbf{q}_{\perp}+\textbf{k}_{\perp})^2+q_{z}^2=q_{\perp}^2+k_{\perp}^2+q_{z}^2+2 \ \textbf{q}_{\perp} \cdot \textbf{k}_{\perp} \notag\\
	&=q^2+k_{\perp}^2+2qk_{\perp}\cos\theta.
	\end{align}
	The integral (\ref{alig:int}) is then transformed as follows:
	\allowdisplaybreaks\begin{align}
	K&=\int dq^{3} \cdot \frac{\epsilon_{B}(q) \epsilon_{B}(a)}{q^{2}a^{2}} [(2-\frac{q_{\perp}^{2}}{q^{2}})(2-\frac{a_{\perp}^{2}}{a^{2}})]& \notag\\
	&=\int dq \int_{-1}^{1}  d\cos\theta \ \frac{\epsilon_{B}(q) \epsilon_{B}(a)}{q^{2}a^{2}} \int_{0}^{2\pi} d\varphi \ [1+\sin^2\theta\cos^2\varphi]&\notag\\
	&\quad \ \left[2-\frac{q^2(1-\sin^2\theta\cos^2\varphi)+k_{\perp}^2+2qk_{\perp}\cos\theta}{q^2+k_{\perp}^2+2qk_{\perp}\cos\theta}\right]& \notag\\
	&=\int dq \int_{-1}^{1}  d\cos\theta \ \frac{\epsilon_{B}(q) \epsilon_{B}(a)}{q^{2}a^{2}}
	\bigg[2+\sin^2\theta& \notag\\
	&\quad \ +\Big(1+\frac{3}{4}\sin^2\theta\Big)\Big(\frac{q^2\sin^2\theta}{q^2+k_{\perp}^2+2qk_{\perp}\cos\theta}\Big)\bigg]\pi& \label{term}
	\end{align}
	We integrate this numerically for values of $k_{\perp}$ between $k=10^{-2} \ \text{and} \ 10^3$. We vary the spectral index $\alpha$ between $\alpha=\frac{1}{2}$ and the Kolmogorov-type spectrum $\alpha=5/3$. The results can be seen in Fig. \ref{bild1} (left). The slope of the declining section is not equal to $\alpha$ but depends on it. It is referred to as the polarisation spectrum slope $\alpha^{*}$. If one plots the energy spectrum slope $\alpha$ against the  polarisation spectrum slope $\alpha^{*}$, one can see that there are two different regimes, for roughly $\alpha > 1$ and $\alpha < 1$ (see Fig. \ref{bild2} (right) ).
	
	It is now necessary to understand the approximate behavior of  our findings in Fig. \ref{bild1} (left)  as well as the significance of the different regimes seen in Fig. \ref{bild2} (right).
	As long as $\alpha$ is large enough the term $\epsilon(q)$ only contributes to the integral within a sphere with radius $q_{0}$ in $q$-space, because there it is mainly constant: $\frac{\epsilon(q)}{q^2}\approx \ \mathrm{const}$. Beyond this sphere it is strongly suppressed by the $q^{-\alpha}$-dependence. In this case the slope is determined by the second term $\epsilon(|\fe{q}+\fe{k}_{\perp}'|)$ in (\ref{term})  which also contributes only within a sphere determined by $q$ around the point $k_{\perp}'$. Inside this sphere, $\epsilon$ is roughly $\epsilon(k_{\perp}')$ because $q$ is small in comparison to $k_{\perp}'$ since we are looking at the case $k_{\perp}'>q_{0}$ (otherwise we get a constant behaviour of the total integral as can be seen in the plots).
	So we get approximately $I\propto \epsilon(0) \epsilon(k_{\perp}')$ which leads to $\log[I] \approx -(\alpha+2)\log[k_{\perp}']+\mathrm{const}$. This is confirmed by the approximations plotted in Fig. \ref{bild1} (left)  and Fig. \ref{bild2} (right). In Fig. \ref{bild1} (left)  we plotted a rough estimate for the integral, where we have just integrated $I\propto \epsilon(0) \epsilon(k_{\perp}')$ inside the $q_{0}$-sphere. In order to match the original integral better, it had to be shifted by a factor of $~1.6$, which is perfectly reasonable considering that the simple sphere is just an approximation for a more complex structure. In Fig. \ref{bild2} (right)  we see that $-\alpha-2$ is indeed a good approximation for $\alpha^{*}$ in the high-$\alpha$ regime.
	The regime where $\alpha<1$ is not really of physical interest because all energy spectra with $\alpha<1$ would lead to the unphysical situation of infinite energies on the smallest scales as the spectrum complies with $k^{-\alpha}$. Therefore we are not interested in the exact behaviour of the integral values below $\alpha=1$.

\section{Other correlation functions}\label{other}

	We now provide the results for the other correlation functions of our observables without repeating the details of the calculations. All functions of first or third order in $\fe{B}$ are omitted because they are obviously zero due to the uneven number of fields appearing in them. These are $\langle \phi(\fe{k}_{\perp}) \rangle_{\fe{B}}$, $\langle \phi(\fe{k}_{\perp}) \phi(\fe{k}'_{\perp}) \phi(\fe{k}''_{\perp}) \rangle_{\fe{B}}$, $\langle I(\fe{k}_{\perp}) \phi(\fe{k}'_{\perp}) \rangle_{\fe{B}}$ and $\langle P(\fe{k}_{\perp}) \phi(\fe{k}'_{\perp}) \rangle_{\fe{B}}$. As stated before, the calculational steps to gain these expressions are fairly similar to the case of $\langle P(\textbf{k}_{\perp}) \cdot P^*(\textbf{k}_{\perp}') \rangle_{B}$. As a matter of fact, most of them are even easier to obtain. More complex numerical integrations are only needed in the cases of $\langle I(\textbf{k}_{\perp}) I(\textbf{k}'_{\perp}) \rangle_{\textbf{B}}$ and $\langle I(\textbf{k}_{\perp}) P(\textbf{k}'_{\perp}) \rangle_{\textbf{B}}$. In the following, the vectors $\fe{r}$ or $\fe{r}'$ shall always denote a combination such as $\fe{x}'-\fe{x}$, to be defined for each correlation function in Appendix \ref{ap:cal}. By $\mathcal{B}(a,b)$ we denote the Beta--function. Furthermore $\fe{u},\fe{v} \ \text{and} \ \fe{w}$ are defined as $\fe{w}=(\fe{k}_{\perp}''',0), \fe{u}=(\fe{k}_{\perp}'',0)$, $\fe{v}=(\fe{k}_{\perp}',0)$ and $\textbf{a}=(-\textbf{q}_{\perp}-\textbf{k}_{\perp}',-q_{z})$. \\
	Here are the results:
		
	\begin{align}
	&\langle I(\textbf{k}_{\perp}) \rangle_{\textbf{B}}&&=2 (2\pi)^2 \delta^2(\textbf{k}_{\perp}) L_{z} M_{N}(0)\notag\\
	& &&=128 \pi^4 \delta^2(\textbf{k}_{\perp}) L_{z} \mathcal{B}\Big(\frac{\beta}{2}+\frac{1}{2},\frac{\alpha}{2}-\frac{1}{2}\Big)\label{al:I} \\
	&\langle P(\textbf{k}_{\perp}) \rangle_{\textbf{B}}&&=0\label{al:P} \\
	&\langle\phi(\textbf{k}_{\perp})\phi(\textbf{k}'_{\perp})\rangle_{\textbf{B}} &&=(2\pi)^2 \delta^2(\textbf{k}_{\perp}+\textbf{k}'_{\perp}) L_{z} \hat{M}_{33}(\textbf{k}'_{\perp},0) \notag\\
	& &&=32 \pi^5 \delta^2(\textbf{k}_{\perp}+\textbf{k}'_{\perp}) L_{z} \epsilon(v)/v^2\label{al:envo} \\
	&\langle I(\textbf{k}_{\perp}) I(\textbf{k}'_{\perp}) \rangle_{\textbf{B}}&&=(2\pi)^{2} \delta^{2}(\fe{k}_{\perp}+\fe{k}_{\perp}') L_{z} \int \frac{dq^3}{(2\pi)^3}  \Big(2\big(\hat{M}_{11}(\fe{q})\hat{M}_{11}(\fe{a}) \notag\\
	& &&\quad +\hat{M}_{22}(\fe{q})\hat{M}_{22}(\fe{a})\big)+4\hat{M}_{21,\mathrm{isym}}(\fe{q})\hat{M}_{21,\mathrm{isym}}(\fe{a})\Big) \notag\\
	& &&=8\pi^2 \delta^{2}(\fe{k}_{\perp}+\fe{k}_{\perp}') L_{z} \int dq^3 \frac{\epsilon(q) \epsilon(a)}{q^2a^2}\Bigg[\bigg(1-\frac{q_{1}^{2}}{q^2}\bigg)\bigg(1-\frac{a_{1}^{2}}{a^2}\bigg) \notag\\
	& &&\quad+\bigg(1-\frac{q_{2}^{2}}{q^2}\bigg)\bigg(1-\frac{a_{2}^{2}}{a^2}\bigg)+2\bigg(\frac{q_{2}q_{1}}{q^2}\bigg)\bigg(\frac{a_{2}a_{1}}{a^2}\bigg)\Bigg] \label{al:IIrs} \\
	&\langle P(\textbf{k}_{\perp}) P^*(\textbf{k}'_{\perp}) \rangle_{\textbf{B}}&&=\frac{8}{2\pi} \delta^{2}(\textbf{k}_{\perp}-\textbf{k}_{\perp}') L_{z} \int dq^{3} \notag\\ 
	& &&\quad \big(M_{11}(\textbf{q})M_{11}(\textbf{a})+M_{22}(\textbf{q})M_{22}(\textbf{a}) \notag\\
	& &&\quad  +M_{22}(\textbf{q})M_{11}(\textbf{a})+M_{22}(\textbf{a})M_{11}(\textbf{q})\big) \notag\\
	& &&=8(2 \pi)^{5} \delta^{2}(\textbf{k}_{\perp}-\textbf{k}_{\perp}') L_{z} \int dq^{3} \cdot \frac{\epsilon_{B}(q) \epsilon_{B}(a)} {q^{2}a^{2}} \notag\\ 
	& &&\quad \Big[\Big(2-\frac{q_{\perp}^{2}}{q^{2}}\Big)\Big(2-\frac{a_{\perp}^{2}}{a^{2}}\Big)\Big] \label{al:PPrs} \\
	&\langle\phi(\textbf{k}_{\perp}) \phi(\textbf{k}'_{\perp}) \phi(\textbf{k}''_{\perp})\phi(\textbf{k}'''_{\perp})  \rangle_{\textbf{B}}&&=(2\pi)^4 \Big[\delta^2(\fe{k}''_{\perp}+\fe{k}'''_{\perp}) \delta^2(\fe{k}_{\perp}+\fe{k}'_{\perp}) \hat{M}_{N}(w)\hat{M}_{N}(v) \notag\\
	& &&\quad +\delta^2(\fe{k}'''_{\perp}+\fe{k}'_{\perp}) \delta^2(\fe{k}_{\perp}''+\fe{k}_{\perp}) \hat{M}_{N}(w)\hat{M}_{N}(u) \notag\\
	& &&\quad +\delta^2(\fe{k}''_{\perp}+\fe{k}'_{\perp}) \delta^2(\fe{k}_{\perp}'''+\fe{k}_{\perp}) \hat{M}_{N}(w)\hat{M}_{N}(u)\Big]\label{al:phphphph} \\ 
	&\langle I(\textbf{k}_{\perp}) P(\textbf{k}'_{\perp}) \rangle_{\textbf{B}}&&=(2\pi)^{2} \delta^{2}(\fe{k}_{\perp}+\fe{k}_{\perp}') L_{z} \int \frac{dq^3}{(2\pi)^3} \bigg[\frac{M_{N}(q) M_{N}(a)}{q^2 a^2}  \notag\\
	& &&\quad \Big[2(q_{2}^2+q_{3}^2)(a_{2}^2+a_{3}^2)+2(q_{1}^2+q_{3}^2)(a_{1}^2+a_{3}^2)\Big]\bigg] \\
	&\langle I(\textbf{k}_{\perp}) \phi(\textbf{k}'_{\perp})\phi(\textbf{k}''_{\perp})\rangle_{\textbf{B}}&&=L^{2} (2\pi)^{4} \delta^{2}(\fe{k}'_{\perp}+\fe{k}''_{\perp}) \delta^{2}(\fe{k}_{\perp})\hat{M}_{N}(u)2M_{N}(0) \notag\\
	& &&\quad -2 L_{z} (2\pi)^2 \delta^{2}(\fe{k}_{\perp}+\fe{k}'_{\perp}+\fe{k}''_{\perp}) \hat{H}(u)\hat{H}(v)/uv \notag\\ 
	& &&\quad \Big(u_{1}v_{1}+u_{2}v_{2}\Big) \label{al:IPP} \\
	&\langle P(\textbf{k}_{\perp}) \phi(\textbf{k}'_{\perp})\phi(\textbf{k}''_{\perp})\rangle_{\textbf{B}}&&=2 L_{z} (2\pi)^2 \delta^{2}(\fe{k}_{\perp}+\fe{k}'_{\perp}+\fe{k}''_{\perp}) \notag\\
	& &&\quad \ \hat{H}(u)\hat{H}(v)/uv \ \Big(\big(u_{1}v_{1}-u_{2}v_{2}\big)+i\big(u_{1}v_{2}+u_{2}v_{1}\big)\Big)\label{al:Ppp}
	\end{align}
	Many of the results are, as expected, providing no surprises.
	The mean total intensity (\ref{al:I}) is given, in principle, by the energy density of the magnetic field whereas the mean polarized intensity (\ref{al:P}) should be zero due to the isotropy of the problem.
	The correlation function (\ref{al:envo}) was already evaluated by \citet{2003A&A...401..835E} and also by \citet{2009ApJ...705L..90C}, it depends on the $k_{z}=0$ plane of $\hat{M}_{zz}$.
	The fourth order quantities (\ref{al:IIrs}), (\ref{al:PPrs}) and (\ref{al:phphphph}) are more complex, but nevertheless, only correlated combinations of the second order quantities. They can be used to monitor the validity of the assumption of Gaussianity and isotropy. Non-Gaussianity or anisotropy in magnetic field statistics would lead to a deviation from this form, which by comparison to $\langle\phi(\textbf{k}_{\perp})\phi(\textbf{k}'_{\perp})\rangle_{\textbf{B}}$ can be detected.
	
\begin{figure}[t]
	\centering
			\includegraphics[width=0.8\textwidth]{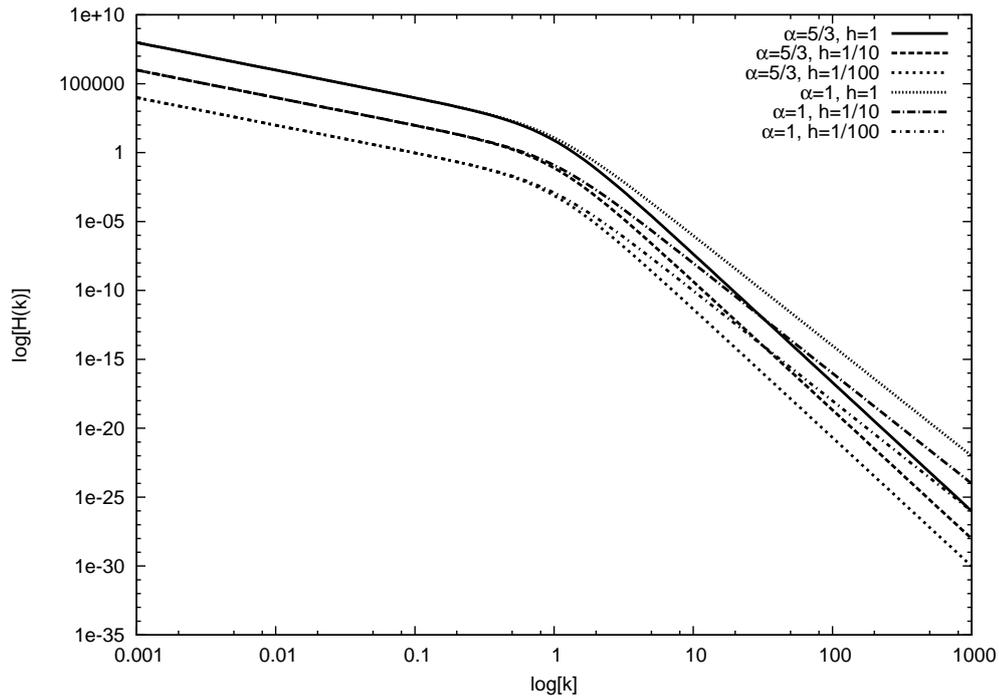}
			\label{PPhPh_a53}
	\caption{Helical spectra of $\langle I(\textbf{k}_{\perp}) \phi(\textbf{k}'_{\perp})\phi(\textbf{k}''_{\perp})\rangle_{\textbf{B}}$ and $\langle P(\textbf{k}_{\perp}) \phi(\textbf{k}'_{\perp})\phi(\textbf{k}''_{\perp})\rangle_{\textbf{B}}$ for $\alpha=1$ and $\alpha=5/3$, $\fe{k}_{\perp}=0$ and different values of $h(k)$ and under further assumptions for which both functions take on the same analytical form.} 
	\label{helicpic}
	\end{figure}

	By far the most interesting results are of course (\ref{al:IPP}) and (\ref{al:Ppp}) as they contain a direct dependence on the helical power spectrum $\hat{H}(k)$. Both are plotted in Fig. (\ref{helicpic}) at the same time, but only for the case where $\fe{k}_{\perp}=0$. We also restrict us to $\fe{k}=k \ \fe{e}_{y}$ for (\ref{al:Ppp}) without loss of generality. This reduces both correlation functions to $2(2\pi)^2\hat{H}^{2}(k)$. Actually plotted is only $\hat{H}^{2}(k)$. To represent the helical spectrum $\hat{H}^{2}(k)$ graphically, we factor out the energy spectrum $\epsilon_{B}(k)$ thus leaving us a function $h(k)$, that parameterises the plots (see Sec. (\ref{M})).
	
	The exact results (\ref{al:IPP}) and (\ref{al:Ppp}) can be understood from the physical point of view. In both correlation functions we have a part connected to the magnetic field component that lies in the surface perpendicular to the line of sight $I(\fe{k}_{\perp})$ and $P(\textbf{k}_{\perp})$ due to the polarisation properties of synchrotron emission. There is also a part which depends on the line-of-sight component $\phi(\textbf{k}_{\perp})\phi(\textbf{k}'_{\perp})$ due to Faraday rotation. If we consider the corresponding delta functions, we see that in cases, where we have $\delta^{2}(\fe{k}'_{\perp}+\fe{k}''_{\perp}) \delta^{2}(\fe{k}_{\perp})$ the two parts are somewhat uncorrelated as the related term is more or less the product of $\langle I(\textbf{k}_{\perp}) \rangle_{\textbf{B}}$ and $\langle\phi(\textbf{k}_{\perp})\phi(\textbf{k}'_{\perp})\rangle_{\textbf{B}}$. This is reflected in the absence of any helicity dependence. In contrast, if we examine the parts with $\delta^{2}(\fe{k}_{\perp}+\fe{k}'_{\perp}+\fe{k}''_{\perp})$ there is a type of mixing between the different observables due to the mutual dependence of the three vectors $\fe{k}_{\perp},\fe{k}'_{\perp},\fe{k}''_{\perp}$ through the delta function. Accordingly, there is a dependence on helicity. More about the physical interpretation of these results can be found in the following section.

	Please note that information on the overall sign of the helicity cannot be obtained using this method becaus $\hat{H}(k)$ only appears quadratically.

\section[The LITMUS test]{The \textit{LITMUS} test}\label{acidtest}

\begin{figure*} [t]
	\centering
	\includegraphics[width=0.8\textwidth]{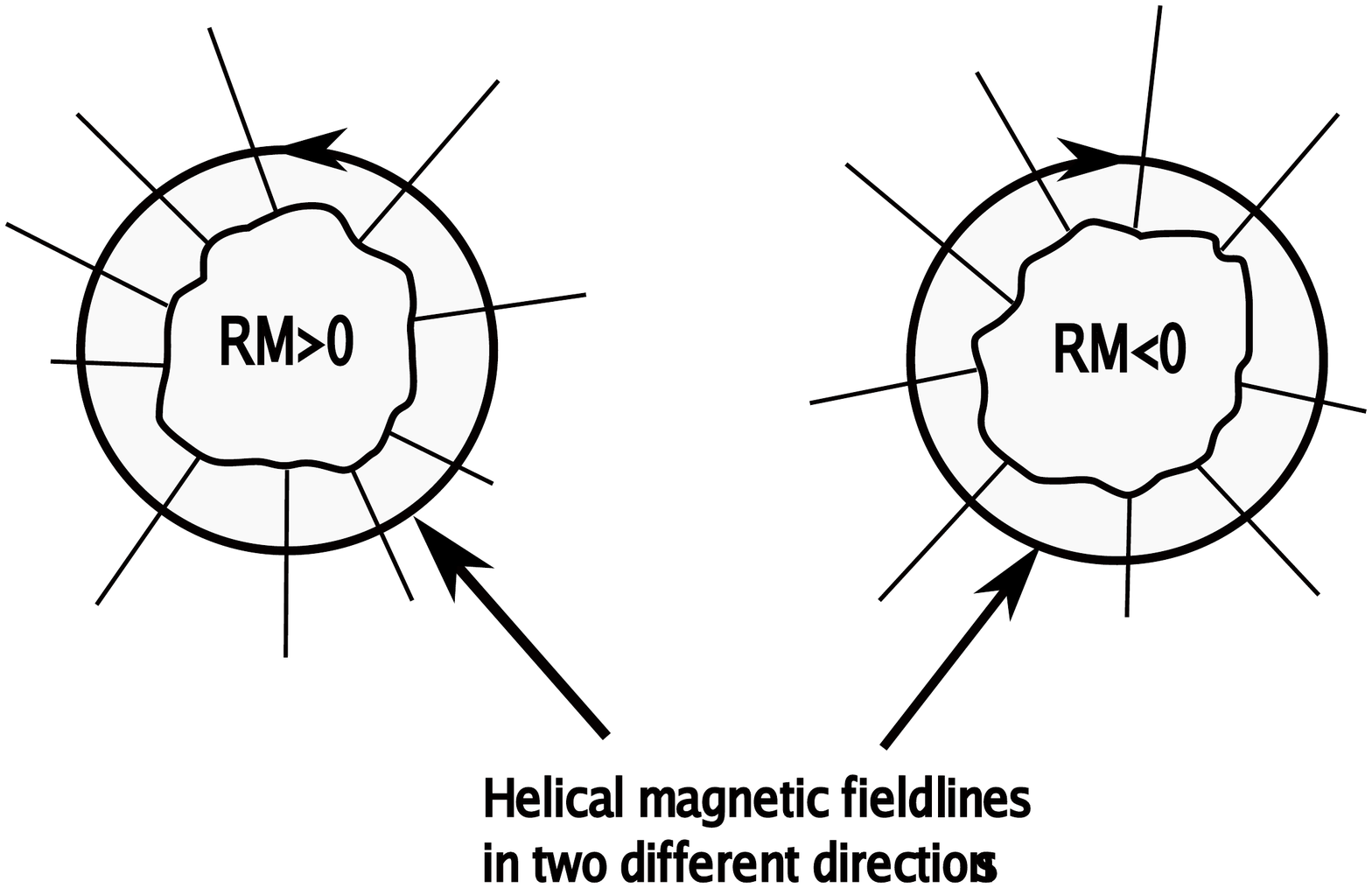}
	\caption{Schematic picture of correlated structures in combined polarisation and RM maps that can give rise to a non-zero correlation function $\langle P(\textbf{k}_{\perp}) \phi(\textbf{k}'_{\perp})\phi(\textbf{k}''_{\perp})\rangle_{\textbf{B}}$.}
	\label{Sonne}
	\label{grad}
	\end{figure*}

	For the correlation function (\ref{al:Ppp}), a strikingly intuitive picture can be found to explain the result. Let us take a look at Fig. (\ref{Sonne}) where we imagine the line of sight to be directly aligned with the axis of a magnetic helix. In a combined polarisation and Faraday depth map, we should see a central region with nonzero faraday depth $\phi$ and around it a radial polarisation pattern. These are correlated structures, that make (\ref{al:Ppp}) nonzero for helical magnetic fields. They would vanish with the field becoming nonhelical. However, we clearly have two possibilities for the direction of the magnetic field going around and therefore could get positive or negative $\phi$ respectively. Thus, these correlated structures can only be seen in $\langle P(\textbf{k}_{\perp})\phi(\textbf{k}'_{\perp})\phi(\textbf{k}''_{\perp})\rangle_{\textbf{B}}$ and not in $\langle P(\textbf{k}_{\perp}) \phi(\textbf{k}'_{\perp})\rangle_{\textbf{B}}$, where the single dependence on $\phi$ would induce the positive and negative parts to cancel out over averaging. This is confirmed, as $\langle P(\textbf{k}_{\perp}) \phi(\textbf{k}'_{\perp})\rangle_{\textbf{B}}$ becomes zero due to the odd number of functional derivatives.

	Guided by this picture, the \textit{LITMUS} test (\fe{L}ocal \fe{I}nference \fe{T}est for \fe{M}agnetic fields which \fe{U}ncovers helice\fe{S}) was developed, a small and simple test that could be easily used to probe real data for helicity. 

	The \textit{LITMUS} test was not constructed to produce quantitative measurements of the helicity spectra, but to provide a fast and qualitative test for the presence of helicity. 

	\subsection{The basic idea}
	We take a closer look at one of the aforementioned patterns of polarisation and Faraday rotation (see Fig. \ref{Sonne}). In the ideal, helical case, the gradient $\fe{G}=\nabla \phi$ of $\phi$ points either to the center or out of the center of the region with Faraday rotation and should therefore be perfectly aligned with the polarisation. The polarised intensity $P$ is a complex number representing a spin 2 field. To compare $P$ with $\fe{G}$, we just transform $\fe{G}$ from a two-dimensional vector into a complex number $\mathcal{G}$ in the same representation of ``directionless vectors'': 
	\begin{align}
	&\mathcal{G}=|\fe{G}|^2 \exp[2i\alpha]  \, \quad \text{with} \ \alpha=\arctan\frac{G_{y}}{G_{x}}.
	\end{align}
	By doing so, we loose the information on the direction in which the gradient $\fe{G}$ is pointing. The quadratic dependence of $\mathcal{G}$ on $|\fe{G}|$ accounts for a normalisation that will become clear in the following. 
	
	We now want to construct a test using $\mathcal{G}$ and $P$ that is sensitive to the presence of magnetic helicity but can be performed easily on a real dataset.
	Our previous considerations suggest to simply use the fact that $\mathcal{G}$ and $P$ should be parallel for helical fields. The test then just consists of multiplying $\mathcal{G}$ with $P^*$ for every pixel of a given map of $\phi^2$ and $P$. This complex scalar product produces different results for different orientations of $\mathcal{G}$ and $P$ in the complex plane. If the gradient and the polarisation are parallel ($\chi$ and $\alpha$ differ by a multiple of $\pi$) the result is real and positive. If they are perpendicular ($\chi$ and $\alpha$ differ by an odd multiple of $\pi/2$) the result is real and negative. For any orientations in between, the result will take on complex values.

	We now can state the \textit{LITMUS} test. In the presence of helical fields, the average of the scalar product $\mathcal{G}^*P$ over all pixels of a $\mathcal{G}^*P$-map should have a real value significantly larger than 0. Whereas in the case of non-helical fields, the alignment of $\mathcal{G}$ and $P$ should be changing randomly from pixel to pixel so that we would expect the average over $\mathcal{G}^*P$ to be zero.
	In short mathematical notation this is stated as 
	\begin{align}
	&\Big\langle \mathcal{G}^*P \ \Big\rangle_{\text{helicity}} > 0 \quad \text{ and real}, \label{con1} \\
	&\Big\langle \mathcal{G}^*P \ \Big\rangle_{\text{no helicity}} = 0 \label{con2},
	\end{align}
	where the ensemble average over helical or nonhelical fields is in practice replaced by an average over all pixels of a $\mathcal{G}^*P$-map.

	It can be shown that our intuition was right and that this test actually depends directly on $\langle P(\textbf{k})\phi(\textbf{k}')\phi(\textbf{k}'')\rangle$ which was our intial starting point.
	We begin by writing out the full condition for the \textit{LITMUS} test:
	\begin{equation}
	\Bigg\langle \Big(\mathcal{G}^*(\fe{x})P(\fe{x}) \Big) \Bigg\rangle=\Bigg\langle \ \Bigg[\Bigg(\frac{\partial \phi(\fe{x})}{\partial x}\Bigg)^2+\Bigg(\frac{\partial \phi(\fe{x})}{\partial y}\Bigg)^2\Bigg] \ \exp\Big[-2i\arctan{G_{y}/G_{x}}\Big] P(\fe{x}) \Bigg\rangle.
	\end{equation}
	Using trigonometrical theorems it is easy to show that
	\begin{equation}
	\Big(G_{x}^2+G_{y}^2\Big) \ \exp\Big[-2i\arctan{G_{y}/G_{x}}\Big]=\Big(G_{x}-iG_{y}\Big)^2 .
	\end{equation}
	We now apply a Fourier transformation to our observables $P(x)$ and $\phi(x)$ and rewrite the whole expression:
	\begin{align}
	\Big\langle \mathcal{G}^*(\fe{x})P(\fe{x}) \Big\rangle&=\Bigg\langle \Bigg[\frac{\partial \phi(\fe{x})}{\partial x_{1}}-i\frac{\partial \phi(\fe{x})}{\partial x_{2}}\Bigg]^2 P(\fe{x})\Bigg\rangle \notag\\
	&=\Bigg\langle \Bigg(\bigg[\frac{\partial}{\partial x_{1}}-i\frac{\partial}{\partial x_{2}}\bigg] \ \phi(\fe{x})\Bigg) \ \Bigg(\bigg[\frac{\partial}{\partial x_{1}}-i\frac{\partial}{\partial x_{2}}\bigg] \ \phi(\fe{x})\Bigg) \ P(\fe{x}) \Bigg\rangle \notag\\
	&=\Bigg\langle \int \frac{dk^2}{(2\pi)^2} \int \frac{dk^{2'}}{(2\pi)^2} \int \frac{dk^{2''}}{(2\pi)^2} \ \Big[k_{1}'-ik_{2}'\Big] \Big[k_{1}''-ik_{2}''\Big] \exp\big[ i\fe{x}\big(\fe{k}+\fe{k}'+\fe{k}''\big)\big] \notag\\
	&\quad P\big(\fe{k}\big) \phi\big(\fe{k}'\big) \phi\big(\fe{k}''\big)\Bigg\rangle \notag\\
	&=\int \frac{dk^2}{(2\pi)^2} ...\int \frac{dk^{2''}}{(2\pi)^2}  \ \Big[k_{1}'-ik_{2}'\Big] \Big[k_{1}''-ik_{2}''\Big] \ \exp\big[ i\fe{x}\big(\fe{k}+\fe{k}'+\fe{k}''\big)\big] \notag\\
	&\quad \Bigg\langle P\big(\fe{k}\big) \phi\big(\fe{k}'\big) \phi\big(\fe{k}''\big)\Bigg\rangle
	\end{align}
	We assumed again that the average over all pixels of a $\mathcal{G}^*P$-map is equivalent to an ensemble average over the magnetic field statistics. 

	 Now we insert our result (\ref{al:Ppp}) for $\langle P(\textbf{k})\phi(\textbf{k}')\phi(\textbf{k}'')\rangle$ and see the dependence of $\Big\langle \mathcal{G}(\fe{x})P^*(\fe{x}) \Big\rangle$ on the helical spectra $\hat{H}(k')\hat{H}(k'')$:
	\begin{align}
	\Big\langle \mathcal{G}(\fe{x})P^*(\fe{x}) \Big\rangle&=2 L_{z} (2\pi)^2 \int \frac{dk^2}{(2\pi)^2} ...\int \frac{dk^{2''}}{(2\pi)^2} \ \delta^{2}(\fe{k}+\fe{k}'+\fe{k}'') \Big[k_{1}'-ik_{2}'\Big] \Big[k_{1}''-ik_{2}''\Big] \notag\\
	&\quad \exp\big[ i\fe{x}\big(\fe{k}+\fe{k}'+\fe{k}''\big)\big] \Big[\big(k_{1}'k_{1}''-k_{2}'k_{2}''\big)+i(k_{1}'k_{2}''+k_{2}'k_{1}''\big)\Big] \ \frac{\hat{H}(k')\hat{H}(k'')}{k'k''} \notag\\ \label{xzero}
	&=2 L_{z} (2\pi)^2 \int \frac{dk^{2'}}{(2\pi)^2} \int \frac{dk^{2''}}{(2\pi)^2} \Big[k_{1}'-ik_{2}'\Big] \Big[k_{1}''-ik_{2}''\Big]\Big[\big(k_{1}'k_{1}''-k_{2}'k_{2}''\big)+i(k_{1}'k_{2}''+k_{2}'k_{1}''\big)\Big] \notag\\
	&\quad \frac{\hat{H}(k')\hat{H}(k'')}{k'k''} \\ 
	&=2 L_{z} (2\pi)^2 \int \frac{dk^{2'}}{(2\pi)^2} \int \frac{dk^{2''}}{(2\pi)^2} \ \Big[k_{1}^{2'}+k_{2}^{2'}\Big] \ \Big[k_{1}^{2''}+k_{2}^{2''}\Big] \ \frac{\hat{H}(k')\hat{H}(k'')}{k'k''} \notag\\
	&=2 L_{z} (2\pi)^2 \int \frac{dk^{2'}}{(2\pi)^2} \int \frac{dk^{2''}}{(2\pi)^2} \ k' \ k'' \ \hat{H}(k')\hat{H}(k'') \notag\\
	&=2 L_{z} \ \Bigg[\int_{0}^{\infty} dk \ k^2 \hat{H}(k) \ \Bigg]^2=2 L_{z} \pi^4 \ \Bigg[\int_{0}^{\infty} dk \  \frac{\epsilon_{H}(k)}{k} \ \Bigg]^2 \label{finalest}
	\end{align}
	In (\ref{xzero}) we assumed w. l. of g. that $\fe{x}=0$ and in the last line we used (\ref{epsH}) to substitute $\epsilon_{H}$.
	We see that $\Big\langle \mathcal{G}(\fe{x})P^*(\fe{x}) \Big\rangle$ is a direct and clear estimator which measures the square of the k-space integrated helicity spectrum, giving more weight to the large scales. Therefore, we conclude that, except maybe from pathological or fine tuned situations in which the k-weighted helicities at different k-scales cancel each other (since $\hat{H}(k)$ might change sign), we can expect the \textit{LITMUS} test to be able to reveal the presence of magnetic helicity. This is demonstrated in numerical tests on simulated data conducted by \citet{Niels}. This work also contains an application on real data of our own galaxy and a thorough analysis thereof.\footnote{Unfortunately, it seems that a straightforward application of this test can be hampered by a too large spatial variance in the elctron density, which we assumed to be constant in this work.}

	Finally, we can state that, in principle, a given dataset can be tested for helicity. If $P$ and $\phi$ are available, a test using the conditions (\ref{con1}) and (\ref{con2}) can easily be implemented due to the usage of local quantities only, avoiding any data transformation to Fourier space and the complication finite window functions thereby would introduce.

\section{Conclusions}\label{con}
	We have shown how statistical properties of turbulent cosmic magnetic fields can imprint on the statistics of certain radio observables. Our analysis involved the total intensity $I(\fe{k}_{\perp})$ and the polarised intensity $P(\textbf{k}_{\perp})$ coming from radio synchrotron emission out of a volume as well as the Farady depth $\phi(\textbf{k}_{\perp})$ of background sources seen through the same volume. The first two depend on $B_{\perp}$, the magnetic field component lying in the surface perpendicular to the line of sight. In contrast, $\phi(\textbf{k}_{\perp})$ depends on $B_{\parallel}$, the component parallel to the line of sight. Whenever this set of observables is available, we can examine all three components of the magnetic field.
	
	With regard to these observables, we evaluated a complete set of cross-correlation functions up to fourth order in the magnetic field and presented simple analytical equations in Fourier space depending on the field's energy spectra. We demonstrated that two correlation functions of our set, namely $\langle P(\textbf{k}_{\perp}) \phi(\textbf{k}'_{\perp})\phi(\textbf{k}''_{\perp})\rangle_{\textbf{B}}$ and $\langle I(\textbf{k}_{\perp}) \phi(\textbf{k}'_{\perp})\phi(\textbf{k}''_{\perp})\rangle_{\textbf{B}}$, explicitly depend on the helical spectra of the turbulent field. The first one depends solely on the helical parts and becomes zero for non-helical fields. 
	
	This finding offers a new way for measuring the helicity of magnetic fields and thereby for testing existing mean field dynamo theories involving helicity. Measuring these correlation functions in real data will permit the study of helicity spectra up to their overall sign. If (\ref{al:IPP}) or (\ref{al:Ppp}) provide a non-zero result with statistical significance for $\fe{k}_{\perp}\neq0$ , this is direct evidence for helicity in the magnetic field.
	
	Furthermore, we presented the \textit{LITMUS} test, a simple procedure to be applied to data which probes for helicity. The \textit{LITMUS} test is easy to apply since it can be fully computed in real space. It provides the square of the k-weighted k-space integrated helicity. First results of an application of the test to real data and to numerical tests using simulated helical and non-helical fields can be found in \citet{Niels}.
	
	Our general formalism permits the construction of further tests which can probe helicity on invidual k-scales. However, before such tests can be applied to real data, suitable observational configurations and further theoretical development in order to alleviate the simplifications and assumptions made beforehand are required.
	
	Subsequent work should therefore follow two directions: To find observations that match our assumptions best and to extend our calculations to be able to cope with more complex observational situations. Advancements should include
	\begin{itemize}
	\item more realistic non-Gaussian components of the magnetic field statistics
	\item the removal of statistical homogeneity as an overall simplification
	\item spatially varying electron densities instead of assuming them as constant (see Appendix \ref{sync})
	\item the possibility for preciser values for the spectral index of the cosmic ray electron density than the choice $p=3$ (see Appendix \ref{sync})
	\item calculations without the restrictions imposed by observed space being large leading to the approximation $\int_{\text{source}}^{\text{observer}}dz\approx \int_{-\infty}^{\infty}dz$ (see Sec. \ref{sec:gen})
	\item the introduction of a window function formalism
	\item and finally developing an approach for dealing with intrinsic Faraday rotation which modifies the polarisation at long wavelengths, a topic which has been neglected here (see Appendix \ref{sync}).
	\end{itemize}
	Before becoming more deeply involved in discussions on possible advances in the future, we should first think about which observations could be applicable to our approach in the present form.
	Observations required by our analysis have to come from a polarized radio-synchrotron source with background Faraday-rotation. A suitable target could probably be found in the interstellar medium (ISM) within our own galaxy, of which we have some established knowledge in respect of large scale fields. With regard to the latter point, it would be advantageous to choose a region in which the magnetic helicity flows are expected to be found in accordance with mean field theory. Furthermore, the polarisation data has to be taken at high frequencies to be Faraday rotation free. For that, the upcoming Planck polarisation data of our Galaxy will be ideal, since it is at short wavelength, has high resolution and accuracy and is also full sky. Compilations of RM measurements of background sources seen through our galaxy already exist \citep{2007ASPC..365..242H,2007ApJ...663..258B,2009ApJ...702.1230T}.
	
	Attractive extragalactic objects to be investigated for magnetic field statistics are the lobes of radio galaxies, whose intensity and polarisation statistics can be constructed. However, no Faraday rotation could be detected yet through their lobes. For the radio jets, this is different and there, helicity can be probed and is yet actually expected to be present \citep{2003A&A...401..499E,2005ASPC..345..264G,2008ASPC..386..494M}.
	
	 Galaxy clusters are probably not well suited for our approach, although they host large scale magnetic fields. The high degree of intrinsic Faraday rotation usually found there erases polarisation of the cluster radio halo emission at the synchrotron frequencies we observe them.

	However, handling intrinsic Faraday-rotation analytically is a considerable challenge. We would need to include an extra exponential factor for the rotation in $P(\fe{x}_{\perp})$:
	\begin{align}
	P(\fe{x}_{\perp})=\int dz \cdot [B_{1}(\fe{x})+iB_{2}(\fe{x})]^{2} \cdot \exp[2i\phi(\fe{x}_{\perp})\lambda^{2}].
	\end{align}
	As $\phi(\fe{x}_{\perp})$ itself contains an integration over $dz$, the extra exponential factor couples all positions along the line of sight. Thus, the exponential becomes so complicated that it has to be approximated in a suitable way. This seems to spoil a purely analytical approach. Nevertheless, the inclusion of intrinsic Faraday rotation is one of the next important challenges as it would enable our approach to be applied to many sources excluded until now, such as galaxy clusters. 
	
	To resume further our basic discussion, there is more to consider. From a technical aspect, the first problem to  be tackled for tests, which aim to measure detailed magnetic energy and helicity spectra, is the inclusion of a realistic window function into the formalism. A window function is set by the observations but also incorporates variation of the signal due to changing relativistic electron density and magnetic field strength. It scales with the electron density $n_{e}(\fe{x})$ for $\phi(\textbf{k}_{\perp})$, with the cosmic ray electron density $n_{cre}(\fe{x})$ for $I(\fe{k}_{\perp})$ and $P(\textbf{k}_{\perp})$ and with the average magnetic field profile. The two former densities have to be taken from independent observations (e.g. free-free emission), the latter has to be guessed depending on the source and relying on prior knowledge. It is clear that introducing a window function will make calculation and integration more complex. In principle, however, there is no basic restriction that could prevent its implementation. Previous attempts to measure magnetic power spectra from observations of magnetic fields heavily affected from window functions have already proven successful, e.g. \citet{2003A&A...401..835E,2005A&A...434...67V,2009arXiv0912.3930K}.
	
	The next aspect to be considered is whether to use Gaussian statistics. Real fields are probably non-Gaussian. Any attempt to model exactly the real situation has to include at least non-Gaussian deviations. This brings additional complexity into our approach as they could not be handled analytically anymore. In (\ref{al:gen}), we would not get rid of the path integral and would need to rely on approximations or perturbative approaches, such as in field theory. 

	At the beginning, our primary goal was, firstly, to find simple analytical relations between the statistics of radio observables and magnetic fields and, secondly, to prove conceptually that with such an approach, it is in principle possible to extract information about the helical part of the magnetic field. To show this, we therefore choose to start with Gaussian statistics, the simplest configuration possible, which also would allow us to keep our calculations analytical while higher order statistics always could be incorporated later in a perturbative expansion around the Gaussian case. We do not claim that this assumption is sufficient to reproduce exact results in accordance with the high complexity of real nature. However, we do believe that while the grade of Gaussianity in our statistics might determine the strength with which the statistics of observables depends on magnetic properties like helicity, the assumption of Gaussianity is not essential for the dependence itself to occur.  Thus we believe we can, in principle, decide whether data contains signatures of helical magnetic fields or not and that the simplification of using Gaussian fields in our calculations is sufficient to achieve the goal of showing how helicity and other magnetic properties can be detected. This of course needs to be shown e.g. using mock data from numerical simulations, but this is left for further studies. 
	
	The reminder of our assumptions and simplifications only represent minor problems. Deviations from the cosmic ray electrons spectral index $p=3$ could be included in the form of correction terms. To assume that the distance between observer and source is very large (and therefore the line of sight projection is parallel) is most of the time a fairly well approximation given the vast distances we encounter on cosmic scales, but it might break down if we analyse the large scale magnetic field directly in front of us. For the \textit{LITMUS} test, the work of \citet{Niels} shows that it is still applicable in such a situation. However the varying $n_{e}$ seems to be a more severe problem. The assumption of statistical homogeneity is widely used in the literature and proved appropriate for similar problems in the past.
	
	The outcome of our study will hopefully contribute to new findings on cosmic magnetic fields. It establishes a new, structured and definite way to measure magnetic helicity and it offers a new option to test cosmic dynamo theories. The presented \textit{LITMUS} test might be able to easily probe data for helicity. However, first results by \citet{Niels} show us that further development is necessary for an actual helicity-sensitive implementation given the complication in realistic observational situations, especially the varying electron density. Last but not least, we present a complete range of correlation functions of radio observables in elegant, simple forms, which are easy to evaluate and, in principle, can be compared to real data. In fact, some of the correlation functions may be of interest in themselves, disregarding the topic of helicity.

\begin{acknowledgements}
This research was performed in the framework of the DFG Forschergruppe 
1254  "Magnetisation of Interstellar and Intergalactic Media: The Prospects 
of Low-Frequency Radio Observations". The idea for this work emerged from the 
very stimulating discussion with Rodion Stepanov during his visit to Germany, 
which was supported by the DFG--RFBR grant 08-02-92881. We like to thank Cornelius Weig, Niels Oppermann, Georg Robbers and an anonymous referee for helpful discussions, and careful reading of our manuscripts.
\end{acknowledgements}

\bibliographystyle{aa}
\bibliography{references.bib}

\begin{thebibliography}{34}
\expandafter\ifx\csname natexlab\endcsname\relax\def\natexlab#1{#1}\fi

\bibitem[{Amsler {et~al.}(2008)Amsler, Doser, Antonelli, Asner, Babu, Baer,
  Band, Barnett, Bergren, Beringer, Bernardi, Bertl, Bichsel, Biebel, Bloch,
  Blucher, Blusk, Cahn, Carena, Caso, Ceccucci, Chakraborty, Chen, Chivukula,
  Cowan, Dahl, D'Ambrosio, Damour, De~Gouvêa, DeGrand, Dobrescu, Drees,
  Edwards, Eidelman, Elvira, Erler, Ezhela, Feng, Fetscher, Fields, Foster,
  Gaisser, Garren, Gerber, Gerbier, Gherghetta, Giudice, Goodman, Grab,
  Gritsan, Grivaz, Groom, Grunewald, Gurtu, Gutsche, Haber, Hagiwara, Hagmann,
  Hayes, Hernández-Rey, Hikasa, Hinchliffe, Höcker, Huston, Igo-Kemenes,
  Jackson, Johnson, Junk, Karlen, Kayser, Kirkby, Klein, Knowles, Kolda,
  Kowalewski, Kreitz, Krusche, Kuyanov, Kwon, Lahav, Langacker, Liddle, Ligeti,
  Lin, Liss, Littenberg, Liu, Lugovsky, Lugovsky, Mahlke, Mangano, Mannel,
  Manohar, Marciano, Martin, Masoni, Milstead, Miquel, Mönig, Murayama,
  Nakamura, Narain, Nason, Navas, Nevski, Nir, Olive, Pape, Patrignani,
  Peacock, Piepke, Punzi, Quadt, Raby, Raffelt, Ratcliff, Renk, Richardson,
  Roesler, Rolli, Romaniouk, Rosenberg, Rosner, Sachrajda, Sakai, Sarkar,
  Sauli, Schneider, Scott, Seligman, Shaevitz, Sjöstrand, Smith, Smoot,
  Spanier, Spieler, Stahl, Stanev, Stone, Sumiyoshi, Tanabashi, Terning, Titov,
  Tkachenko, Törnqvist, Tovey, Trilling, Trippe, Valencia, Van~Bibber, Vincter,
  Vogel, Ward, Watari, Webber, Weiglein, Wells, Whalley, Wheeler, Wohl,
  Wolfenstein, Womersley, Woody, Workman, Yamamoto, Yao, Zenin, Zhang, Zhu,
  Zyla, Harper, Lugovsky, \& Schaffner}]{Amsler:1124989}
Amsler, C., Doser, M., Antonelli, M., {et~al.} 2008, Phys. Lett. B, 667, 1

\bibitem[{{Bracewell}(2000)}]{2000fta..book.....B}
{Bracewell}, R.~N. 2000, {The Fourier transform and its applications}, ed.
  {Bracewell, R.~N.}

\bibitem[{{Brandenburg}(2009)}]{2009PPCF...51l4043B}
{Brandenburg}, A. 2009, Plasma Physics and Controlled Fusion, 51, 124043

\bibitem[{{Brandenburg} \&
  {Subramanian}(2005{\natexlab{a}})}]{2005PhR...417....1B}
{Brandenburg}, A. \& {Subramanian}, K. 2005{\natexlab{a}}, \physrep, 417, 1

\bibitem[{{Brandenburg} \&
  {Subramanian}(2005{\natexlab{b}})}]{2005AN....326..400B}
{Brandenburg}, A. \& {Subramanian}, K. 2005{\natexlab{b}}, Astronomische
  Nachrichten, 326, 400

\bibitem[{{Brown} {et~al.}(2007){Brown}, {Haverkorn}, {Gaensler}, {Taylor},
  {Bizunok}, {McClure-Griffiths}, {Dickey}, \& {Green}}]{2007ApJ...663..258B}
{Brown}, J.~C., {Haverkorn}, M., {Gaensler}, B.~M., {et~al.} 2007, \apj, 663,
  258

\bibitem[{{Cho} \& {Ryu}(2009)}]{2009ApJ...705L..90C}
{Cho}, J. \& {Ryu}, D. 2009, \apjl, 705, L90

\bibitem[{{Eilek}(1989{\natexlab{a}})}]{1989AJ.....98..244E}
{Eilek}, J.~A. 1989{\natexlab{a}}, \aj, 98, 244

\bibitem[{{Eilek}(1989{\natexlab{b}})}]{1989AJ.....98..256E}
{Eilek}, J.~A. 1989{\natexlab{b}}, \aj, 98, 256

\bibitem[{{En{\ss}lin}(2003)}]{2003A&A...401..499E}
{En{\ss}lin}, T.~A. 2003, \aap, 401, 499

\bibitem[{{En{\ss}lin} \& {Biermann}(1998)}]{1998A&A...330...90E}
{En{\ss}lin}, T.~A. \& {Biermann}, P.~L. 1998, \aap, 330, 90

\bibitem[{{En{\ss}lin} \& {Vogt}(2003)}]{2003A&A...401..835E}
{En{\ss}lin}, T.~A. \& {Vogt}, C. 2003, \aap, 401, 835

\bibitem[{{Gabuzda}(2005)}]{2005ASPC..345..264G}
{Gabuzda}, D.~C. 2005, in Astronomical Society of the Pacific Conference
  Series, Vol. 345, Astronomical Society of the Pacific Conference Series, ed.
  {N.~Kassim, M.~Perez, W.~Junor, \& P.~Henning}, 264--+

\bibitem[{{Haverkorn}(2007)}]{2007ASPC..365..242H}
{Haverkorn}, M. 2007, in Astronomical Society of the Pacific Conference Series,
  Vol. 365, SINS - Small Ionized and Neutral Structures in the Diffuse
  Interstellar Medium, ed. {M.~Haverkorn \& W.~M.~Goss}, 242--+

\bibitem[{{Kahniashvili} \& {Vachaspati}(2006)}]{2006PhRvD..73f3507K}
{Kahniashvili}, T. \& {Vachaspati}, T. 2006, \prd, 73, 063507

\bibitem[{{Kronberg} {et~al.}(2008){Kronberg}, {Bernet}, {Miniati}, {Lilly},
  {Short}, \& {Higdon}}]{2008ApJ...676...70K}
{Kronberg}, P.~P., {Bernet}, M.~L., {Miniati}, F., {et~al.} 2008, \apj, 676, 70

\bibitem[{{Kuchar} \& {En{\ss}lin}(2009)}]{2009arXiv0912.3930K}
{Kuchar}, P. \& {En{\ss}lin}, T.~A. 2009, ArXiv e-prints

\bibitem[{{Mahmud} \& {Gabuzda}(2008)}]{2008ASPC..386..494M}
{Mahmud}, M. \& {Gabuzda}, D.~C. 2008, in Astronomical Society of the Pacific
  Conference Series, Vol. 386, Extragalactic Jets: Theory and Observation from
  Radio to Gamma Ray, ed. {T.~A.~Rector \& D.~S.~De Young}, 494--+

\bibitem[{{Moffatt}(1978)}]{1978mfge.book.....M}
{Moffatt}, H.~K. 1978, {Magnetic field generation in electrically conducting
  fluids}, ed. {Moffatt, H.~K.}

\bibitem[{{Narayan} \& {Medvedev}(2001)}]{2001ApJ...562L.129N}
{Narayan}, R. \& {Medvedev}, M.~V. 2001, \apjl, 562, L129

\bibitem[{{Oppermann} {et~al.}(2010){Oppermann}, {Junklewitz}, {Robbers}, \&
  {En{\ss}lin}}]{Niels}
{Oppermann}, N., {Junklewitz}, H., {Robbers}, G., \& {En{\ss}lin}, T. 2010,
  submitted

\bibitem[{{Peskin} \& {Schroeder}(1995)}]{1995iqft.book.....P}
{Peskin}, M.~E. \& {Schroeder}, D.~V. 1995, {An Introduction to Quantum Field
  Theory}, ed. M.~E. Peskin \& D.~V. Schroeder (Westview Press)

\bibitem[{{Price} {et~al.}(2009){Price}, {Bate}, \&
  {Dobbs}}]{2009RMxAC..36..128P}
{Price}, D.~J., {Bate}, M.~R., \& {Dobbs}, C.~L. 2009, in Revista Mexicana de
  Astronomia y Astrofisica Conference Series, Vol.~36, Revista Mexicana de
  Astronomia y Astrofisica Conference Series, 128--136

\bibitem[{{Rybicki} \& {Lightman}(1979)}]{1979rpa..book.....R}
{Rybicki}, G.~B. \& {Lightman}, A.~P. 1979, {Radiative processes in
  astrophysics}, ed. G.~B. Rybicki \& A.~P. Lightman

\bibitem[{{Shukurov} {et~al.}(2006){Shukurov}, {Sokoloff}, {Subramanian}, \&
  {Brandenburg}}]{2006A&A...448L..33S}
{Shukurov}, A., {Sokoloff}, D., {Subramanian}, K., \& {Brandenburg}, A. 2006,
  \aap, 448, L33

\bibitem[{{Sokoloff}(2007)}]{2007PPCF...49..447S}
{Sokoloff}, D. 2007, Plasma Physics and Controlled Fusion, 49, 447

\bibitem[{{Spangler}(1982)}]{1982ApJ...261..310S}
{Spangler}, S.~R. 1982, \apj, 261, 310

\bibitem[{{Spangler}(1983)}]{1983ApJ...271L..49S}
{Spangler}, S.~R. 1983, \apjl, 271, L49

\bibitem[{{Strong} {et~al.}(2007){Strong}, {Moskalenko}, \&
  {Ptuskin}}]{2007ARNPS..57..285S}
{Strong}, A.~W., {Moskalenko}, I.~V., \& {Ptuskin}, V.~S. 2007, Annual Review
  of Nuclear and Particle Science, 57, 285

\bibitem[{{Subramanian}(2002)}]{2002BASI...30..715S}
{Subramanian}, K. 2002, Bulletin of the Astronomical Society of India, 30, 715

\bibitem[{{Taylor} {et~al.}(2009){Taylor}, {Stil}, \&
  {Sunstrum}}]{2009ApJ...702.1230T}
{Taylor}, A.~R., {Stil}, J.~M., \& {Sunstrum}, C. 2009, \apj, 702, 1230

\bibitem[{{Vogt} \& {En{\ss}lin}(2005)}]{2005A&A...434...67V}
{Vogt}, C. \& {En{\ss}lin}, T.~A. 2005, \aap, 434, 67

\bibitem[{{Volegova} \& {Stepanov}(2010)}]{2010JETPL..90..637V}
{Volegova}, A.~A. \& {Stepanov}, R.~A. 2010, Soviet Journal of Experimental and
  Theoretical Physics Letters, 90, 637

\bibitem[{{Waelkens} {et~al.}(2009){Waelkens}, Schekochihin, \&
  En{\ss}lin}]{2009MNRAS.398.1970W}
{Waelkens}, A.~H., Schekochihin, A.~A., \& En{\ss}lin, T.~A. 2009, \mnras, 398,
  1970

\end{thebibliography}

\begin{appendix}

 \section{All correlation functions}\label{ap:cal}
	This Appendix contains a full list of all evaluated correlation functions and their derivations. This constitutes the core of our work but provides no further inside information on the concepts or helps understanding the findings. Therefore, we have collected the calculations in this Appendix. The details of the calculations are similar to the example of $\langle P(\textbf{k}_{\perp}) \cdot P^*(\textbf{k}_{\perp}') \rangle_{B}$, dealt with some depth in Sec. \ref{sss:PP} and are only  commented on if necessary. All correlation functions are given in Fourier space. The functions with an odd number of fields are equal to zero due to the analogous odd number of derivatives with respect to the generating functional $\fe{J}$. The calculations involving $P(\textbf{k}_{\perp})$ are given in a Faraday-free case, but set up in a way so that we could include  Faraday rotational effects for further analysis later on. Refer to chapter \ref{con} for more information on future perspectives. In the following, the vectors $\fe{r}$ or $\fe{r}'$ shall always denote a combination such as $\fe{x}'-\fe{x}$, to be defined for each correlation function in Appendix \ref{ap:cal}. In the following $\fe{u},\fe{v} \ \text{and} \ \fe{w}$ are defined as $\fe{w}=(\fe{k}_{\perp}''',0), \fe{u}=(\fe{k}_{\perp}'',0)$, $\fe{v}=(\fe{k}_{\perp}',0)$ and $\textbf{a}=(-\textbf{q}_{\perp}-\textbf{k}_{\perp}',-q_{z})$.

		\subsection{Calculation of $\langle\phi(\textbf{k}_{\perp})\rangle_{\textbf{B}}$}
			\begin{align}
			&\langle\phi(\textbf{k}_{\perp})\rangle_{\textbf{B}}=\int dx^3 \exp[i\textbf{k}_{\perp}\textbf{x}_{\perp}] \ \partial_{3}(\textbf{x}) \exp\source|_{\textbf{J}=0}=0
			\end{align}

		\subsection{Calculation of $\langle\phi(\textbf{k}_{\perp}) \phi(\textbf{k}'_{\perp}) \rangle_{\textbf{B}}$}
			\allowdisplaybreaks\begin{align}
			&\langle\phi(\textbf{k}_{\perp})\phi(\textbf{k}'_{\perp})\rangle_{\textbf{B}}=\int dx^3 \int dx'^3 \exp[i\textbf{k}_{\perp}\textbf{x}_{\perp}+i\textbf{k}'_{\perp}\textbf{x}'_{\perp}] \ \partial_{3}(\textbf{x}) \ \partial_{3}(\textbf{x}') \notag\\ 
			&\exp\source|_{\textbf{J}=0} \notag\\
			&=\int dx^3 \int dx'^3 \exp[i\textbf{k}_{\perp}\textbf{x}_{\perp}+i\textbf{k}'_{\perp}\textbf{x}'_{\perp}] \ M_{33}(\textbf{r}) \notag\\
			&=\int dx^3 \int dr^3 \exp[i\textbf{x}_{\perp}(\textbf{k}_{\perp}+\textbf{k}'_{\perp})] \exp[i\textbf{r}_{\perp}\textbf{k}'_{\perp}] \ M_{33}(\textbf{r}) \notag\\
			&=(2\pi)^2 \delta^2(\textbf{k}_{\perp}+\textbf{k}'_{\perp}) L_{z}  \int dr^3 \exp[i\textbf{r}_{\perp}\textbf{k}'_{\perp}] \ M_{33}(\textbf{r}) \notag\\
			&=(2\pi)^2 \delta^2(\textbf{k}_{\perp}+\textbf{k}'_{\perp}) L_{z} \int dr^3 \int \frac{dq^3}{(2\pi)^3} \exp[i\textbf{r}_{\perp}\textbf{k}'_{\perp}] \exp[-i\textbf{r}\textbf{q}] \ \hat{M}_{33}(\textbf{q}) \notag\\ 
			&=(2\pi)^2 \delta^2(\textbf{k}_{\perp}+\textbf{k}'_{\perp}) L_{z}  \int dq^3 \delta^2(\textbf{k}'_{\perp}-\textbf{q}_{\perp}) \ \delta(-q_{z}) \  \hat{M}_{33}(\textbf{q}) \notag\\
			&=(2\pi)^2 \delta^2(\textbf{k}_{\perp}+\textbf{k}'_{\perp}) L_{z}  \hat{M}_{33}(\textbf{k}'_{\perp},0) \notag\\
			&=(2\pi)^2 \delta^2(\textbf{k}_{\perp}+\textbf{k}'_{\perp}) L_{z}  \hat{M}_{N}(v) \notag\\
			&=32 \pi^5 \delta^2(\textbf{k}_{\perp}+\textbf{k}'_{\perp}) L_{z}  \frac{\epsilon(v)}{v^2}
			\end{align}

		\subsection{Calculation of $\langle\phi(\textbf{k}_{\perp}) \phi(\textbf{k}'_{\perp}) \phi(\textbf{k}''_{\perp}) \rangle_{\textbf{B}}$}
			\begin{align}
			&\langle\phi(\textbf{k}_{\perp}) \phi(\textbf{k}'_{\perp}) \phi(\textbf{k}''_{\perp}) \rangle_{\textbf{B}}=a_{0}^3 n_{e}^3 \int dx^3 \int dx^{'3} \int dx^{''3} \exp[i\textbf{k}_{\perp}\textbf{x}_{\perp}+i\textbf{k}'_{\perp}\textbf{x}'_{\perp}+\textbf{k}''_{\perp}\textbf{x}''_{\perp}] \notag\\ &\partial_{3}(\textbf{x})|_{\textbf{J}=0}\ \partial_{3}(\textbf{x}')|_{\textbf{J}=0} \ \partial_{3}(\textbf{x}'') \exp\source|_{\textbf{J}=0}=0
			\end{align}

		\subsection{Calculation of $\langle\phi(\textbf{k}_{\perp}) \phi(\textbf{k}'_{\perp}) \phi(\textbf{k}''_{\perp})\phi(\textbf{k}'''_{\perp}) \rangle_{\textbf{B}}$}
			\begin{align}
			&\langle\phi(\textbf{k}_{\perp}) \phi(\textbf{k}'_{\perp}) \phi(\textbf{k}''_{\perp})\phi(\textbf{k}'''_{\perp}) \rangle_{\textbf{B}}=n_{e}^4 \int dx^3 \int dx'^3 \int dx''^3 \int dx'''^3 \notag\\ &\quad \exp[i\fe{k}_{\perp}\fe{x}_{\perp}+i\fe{k}'_{\perp}\fe{x}'_{\perp}+i\fe{k}''_{\perp}\fe{x}''_{\perp}+i\fe{k}'''_{\perp}\fe{x}'''_{\perp}]\big[\partial_{3}(\fe{x})\partial_{3}(\fe{x}')\partial_{3}(\fe{x}'')\partial_{3}(\fe{x}''')\big] \notag\\
			&\quad \exp\source|_{\fe{J}=0} \notag\\
			&=n_{e}^4 \int dx^3 \int dx'^3 \int dx''^3 \int dx'''^3  \exp[i\fe{k}_{\perp}\fe{x}_{\perp}+i\fe{k}'_{\perp}\fe{x}'_{\perp}+i\fe{k}''_{\perp}\fe{x}''_{\perp}+i\fe{k}'''_{\perp}\fe{x}'''_{\perp}] \notag\\
			&\quad \big[\underbrace{M_{33}(\fe{x}'''-\fe{x}'')M_{33}(\fe{x}'-\fe{x})}_{\mathrm{Part 1}}+\underbrace{M_{33}(\fe{x}'''-\fe{x}')M_{33}(\fe{x}''-\fe{x})}_{\mathrm{Part 2}} \notag\\
			&\quad +\underbrace{M_{33}(\fe{x}'''-\fe{x})M_{33}(\fe{x}''-\fe{x}')}_{\mathrm{Part 3}}\big]
			\end{align}
			\textbf{Part 1} gives with $\fe{x}'''-\fe{x}''=\fe{r}'$ and $\fe{x}'-\fe{x}=\fe{r}$:
			\begin{align}
			\text{\fe{Part 1}}&=n_{e}^4 \int dx^3 \int dx''^3 \int dr^3 \int dr'^3 \exp[i\fe{x}''_{\perp}(\fe{k}''_{\perp}+\fe{k}'''_{\perp})] \exp[i\fe{x}_{\perp}(\fe{k}_{\perp}+\fe{k}'_{\perp})]\notag\\
			&\quad \exp[i\fe{r}'_{\perp}\fe{k}'''_{\perp}] \exp[i\fe{r}_{\perp}\fe{k}'_{\perp}] M_{33}(\fe{r}')M_{33}(\fe{r}) \notag\\
			&=n_{e}^4 (2\pi)^4 \delta^2(\fe{k}''_{\perp}+\fe{k}'''_{\perp}) \delta^2(\fe{k}_{\perp}+\fe{k}'_{\perp}) \int dr^3 \int dr'^3 \int \frac{dq^{3}}{(2\pi)^{3}} \int \frac{dq'^{3}}{(2\pi)^{3}} \notag\\
			&\quad \exp[i\fe{r}_{\perp}(\fe{k}_{\perp}'-\fe{q}_{\perp})] \exp[-ir_{z}q_{z}] \exp[i\fe{r}_{\perp}'(\fe{k}_{\perp}'''-\fe{q}_{\perp}')] \exp[-ir_{z}'q_{z}']\notag\\
			&\quad \hat{M}_{33}(\fe{q}')\hat{M}_{33}(\fe{q}) \notag\\
			&=n_{e}^4 (2\pi)^4 \delta^2(\fe{k}''_{\perp}+\fe{k}'''_{\perp}) \delta^2(\fe{k}_{\perp}+\fe{k}'_{\perp}) \hat{M}_{33}(\fe{k}_{\perp}''',0) \hat{M}_{33}(\fe{k}_{\perp}',0).
			\end{align}
			\textbf{Part 2} and \textbf{Part 3} essentially provide the same if one just adopts the definition of $\fe{r}$ and $\fe{r}'$ as follows \\
			\textbf{Part 2}: $\fe{x}'''-\fe{x}'=\fe{r}'$ and $\fe{x}''-\fe{x}=\fe{r}$, \\
			\textbf{Part 3}: $\fe{x}''-\fe{x}=\fe{r}'$ and $\fe{x}'''-\fe{x}=\fe{r}$. \\
			The final result is then
			\begin{align}
			&\langle\phi(\textbf{k}_{\perp}) \phi(\textbf{k}'_{\perp}) \phi(\textbf{k}''_{\perp})\phi(\textbf{k}'''_{\perp}) \rangle_{\textbf{B}}= n_{e}^4 (2\pi)^4 \Big[\delta^2(\fe{k}''_{\perp}+\fe{k}'''_{\perp}) \delta^2(\fe{k}_{\perp}+\fe{k}'_{\perp}) \notag\\
			&\quad \hat{M}_{33}(\fe{k}_{\perp}''',0) \hat{M}_{33}(\fe{k}_{\perp}',0)+\delta^2(\fe{k}'''_{\perp}+\fe{k}'_{\perp}) \delta^2(\fe{k}_{\perp}''+\fe{k}_{\perp}) \hat{M}_{33}(\fe{k}_{\perp}''',0) \hat{M}_{33}(\fe{k}_{\perp}'',0) \notag\\
			&\quad+\delta^2(\fe{k}''_{\perp}+\fe{k}'_{\perp}) \delta^2(\fe{k}_{\perp}'''+\fe{k}_{\perp}) \hat{M}_{33}(\fe{k}_{\perp}''',0) \hat{M}_{33}(\fe{k}_{\perp}'',0)\Big] \notag\\
			&=n_{e}^4 (2\pi)^4 \Big[\delta^2(\fe{k}''_{\perp}+\fe{k}'''_{\perp}) \delta^2(\fe{k}_{\perp}+\fe{k}'_{\perp}) \hat{M}_{N}(w)\hat{M}_{N}(v)+\delta^2(\fe{k}'''_{\perp}+\fe{k}'_{\perp}) \delta^2(\fe{k}_{\perp}''+\fe{k}_{\perp}) \notag\\
			&\quad \hat{M}_{N}(w)\hat{M}_{N}(u)+\delta^2(\fe{k}''_{\perp}+\fe{k}'_{\perp}) \delta^2(\fe{k}_{\perp}'''+\fe{k}_{\perp}) \hat{M}_{N}(w)\hat{M}_{N}(u)\Big] \label{al:pppp}
			\end{align}

		\subsection{Calculation of $\langle I(\textbf{k}_{\perp}) \rangle_{\textbf{B}}$}
			\allowdisplaybreaks\begin{align}
			\langle I(\textbf{k}_{\perp}) \rangle_{\textbf{B}}&=\int dx^3 \exp[i\textbf{k}_{\perp}\textbf{x}_{\perp}] \left( \partial_{1}^2(\textbf{x})+\partial_{2}^2(\textbf{x})\right) \exp\source \vert_{\textbf{J}=0} \notag\\
			&=\int dx^3 \exp[i\textbf{k}_{\perp}\textbf{x}_{\perp}] \left(M_{11}(\textbf{x},\textbf{x})+M_{22}(\textbf{x},\textbf{x})\right) \notag\\
			&=\int dx^3 \exp[i\textbf{k}_{\perp}\textbf{x}_{\perp}] \left(M_{11}(0)+M_{22}(0)\right) \notag\\
			&=2 M_{N}(0) \int dx^3 \exp[i\textbf{k}_{\perp}\textbf{x}_{\perp}] \notag\\
			&=2 (2\pi)^2 \delta^2(\textbf{k}_{\perp}) L_{z} M_{N}(0) \notag\\
			&=128 \pi^4 \delta^2(\textbf{k}_{\perp}) L_{z} \mathcal{B}\Big(\frac{\beta}{2}+\frac{1}{2},\frac{\alpha}{2}-\frac{1}{2}\Big) 	
			\end{align}
			In the final step, $M_{N}(0)$ was expressed in terms of a Beta-function $\mathcal{B}(a,b)$. This function is assigned a specific value for a given set of spectral indices $\alpha$ and $\beta$.
  
		\subsection{Calculation of $\langle I(\textbf{k}_{\perp}) I(\textbf{k}'_{\perp}) \rangle_{\textbf{B}}$}
			\allowdisplaybreaks\begin{align}
			&\langle I(\textbf{k}_{\perp}) I(\textbf{k}'_{\perp}) \rangle_{\textbf{B}}=\int dx^3 \int dx'^3 \exp \left[i(\textbf{k}_{\perp}\textbf{x}_{\perp}+\textbf{k}'_{\perp}\textbf{x}'_{\perp})\right] \Big(\partial_{1}^2(\textbf{x})+\partial_{2}^2(\textbf{x})\Big) \notag\\
			&\quad \Big(\partial_{1}^2(\textbf{x}')+\partial_{2}^2(\textbf{x}')\Big) \exp\source|_{\textbf{J}=0} \notag\\
			&=\int dx^3 \int dx'^3 \exp \left[i(\textbf{k}_{\perp}\textbf{x}_{\perp}+\textbf{k}'_{\perp}\textbf{x}'_{\perp})\right]
			\Big(\partial_{1}^{2}(\fe{x})\partial_{1}^{2}(\fe{x}')+\partial_{1}^{2}(\fe{x})\partial_{2}^{2}(\fe{x}') \notag\\
			&\quad +\partial_{2}^{2}(\fe{x})\partial_{1}^{2}(\fe{x}')+\partial_{2}^{2}(\fe{x})\partial_{2}^{2}(\fe{x}')\Big) \exp\source|_{\fe{J}=0} \notag\\ 
			&=\int dx^3 \int dx'^3 \exp \left[i(\textbf{k}_{\perp}\textbf{x}_{\perp}+\textbf{k}'_{\perp}\textbf{x}'_{\perp})\right] \Big( \underbrace{\big(M_{11}(0)+M_{22}(0)\big)^2}_{\mathrm{Part 1}} \notag\\
			&\quad +\underbrace{2\big(M_{11}^2(\fe{r})+M_{22}^2(\fe{r})\big)+4M_{21,\mathrm{sym}}^{2}(\fe{r})}_{\mathrm{Part 2}}\Big) \notag\\
			\end{align}
			\textbf{Part 1}\footnote{This is actually the same as the results obtained by computing $\langle I(\fe{k}_{\perp})\rangle_{\fe{B}}^2$}  is not dependent on $\fe{r}$ and can be applied directly without difficulties:
			\begin{align}
			&\int dx^3 \int dx'^3 \exp \left[i(\textbf{k}_{\perp}\textbf{x}_{\perp}+\textbf{k}'_{\perp}\textbf{x}'_{\perp})\right]\big(M_{11}(0)+M_{22}(0)\big)^2 \notag\\
			&=(2\pi)^{4} \delta^{2}(\fe{k}_{\perp}) \delta^{2}(\fe{k}_{\perp}') L_{z}L_{z}'\big(M_{11}(0)+M_{22}(0)\big)^2 \notag\\
			&=(2\pi)^{4} \delta^{2}(\fe{k}_{\perp}) \delta^{2}(\fe{k}_{\perp}') L_{z}L_{z}'4M_{N}^2(0) 
			\end{align}
			In contrast, \textbf{Part 2} needs a little more work using Fourier transformations:
			\allowdisplaybreaks\begin{align}
			&\int dx^3 \int dr^3 \exp\left[i\fe{x}_{\perp}(\fe{k}_{\perp}+\fe{k}_{\perp}')\right] \exp\left[i\fe{k}_{\perp}'\fe{r}_{\perp}\right]
			\Big(2\big(M_{11}^2(\textbf{r})+M_{22}^2(\textbf{r})\big)+4M_{21,\mathrm{sym}}^{2}(\textbf{r})\Big) \notag\\
			&=(2\pi)^{2} \delta^{2}(\fe{k}_{\perp}+\fe{k}_{\perp}') L_{z} \int dr^3  \exp\left[i\fe{k}_{\perp}'\fe{r}_{\perp}\right]\Big(2\big(M_{11}^2(\textbf{r})+M_{22}^2(\textbf{r})\big)  +4M_{21,\mathrm{sym}}^{2}(\textbf{r})\Big) \notag\\
			&=(2\pi)^{2} \delta^{2}(\fe{k}_{\perp}+\fe{k}_{\perp}') L_{z} \int dr^3 \int \frac{dq^3}{(2\pi)^3} \int \frac{dq'^3}{(2\pi)^3} \exp\left[i\fe{k}_{\perp}'\fe{r}_{\perp}\right] \exp\left[i\fe{r}(\fe{q}+\fe{q}')\right] \notag\\
			& \quad \Big(2\big(\hat{M}_{11}(\fe{q})\hat{M}_{11}(\fe{q}')+\hat{M}_{22}(\fe{q})\hat{M}_{22}(\fe{q})'\big)
			+4\hat{M}_{21,\mathrm{sym}}(\fe{q})\hat{M}_{21,\mathrm{sym}}(\fe{q}')\Big) \notag\\
			&=(2\pi)^{2} \delta^{2}(\fe{k}_{\perp}+\fe{k}_{\perp}') L_{z} \int dr^3 \int \frac{dq^3}{(2\pi)^3} \int \frac{dq'^3}{(2\pi)^3} \exp\left[i\fe{r}_{\perp}(\fe{q}_{\perp}+\fe{q}_{\perp}'+\fe{k}'_{\perp})\right]\notag\\ 
			&\quad \exp\left[ir_{z}(q_{z}+q_{z}')\right] \Big(2\big(\hat{M}_{11}(\fe{q})\hat{M}_{11}(\fe{q}')+\hat{M}_{22}(\fe{q})\hat{M}_{22}(\fe{q})'\big) \notag\\ 
			&\quad +4\hat{M}_{21,\mathrm{sym}}(\fe{q})\hat{M}_{21,\mathrm{sym}}(\fe{q}')\Big) \notag\\
			&=(2\pi)^{2} \delta^{2}(\fe{k}_{\perp}+\fe{k}_{\perp}') L_{z} \int \frac{dq^3}{(2\pi)^3} \int \frac{dq'^3}{(2\pi)^3} \ (2\pi)^{3}\delta^{2}(\fe{q}_{\perp}+\fe{q}_{\perp}'+\fe{k}'_{\perp})\notag\\ 
			&\quad \delta(q_{z}+q_{z}') \Big(2\big(\hat{M}_{11}(\fe{q})\hat{M}_{11}(\fe{q}')+\hat{M}_{22}(\fe{q})\hat{M}_{22}(\fe{q})'\big) 
			+4\hat{M}_{21,\mathrm{sym}}(\fe{q})\hat{M}_{21,\mathrm{sym}}(\fe{q}')\Big) \notag\\ 
			&=(2\pi)^{2} \delta^{2}(\fe{k}_{\perp}+\fe{k}_{\perp}') L_{z} \int \frac{dq^3}{(2\pi)^3}  \Big(2\big(\hat{M}_{11}(\fe{q})\hat{M}_{11}(\fe{a}) \notag\\
			&\quad +\hat{M}_{22}(\fe{q})\hat{M}_{22}(\fe{a})\big) +4\hat{M}_{21,\mathrm{sym}}(\fe{q})\hat{M}_{21,\mathrm{sym}}(\fe{a})\Big) \notag\\
			&=8\pi^2 \delta^{2}(\fe{k}_{\perp}+\fe{k}_{\perp}') L_{z} \int dq^3 \frac{\epsilon(q) \epsilon(a)}{q^2a^2}\Bigg[\bigg(1-\frac{q_{1}^{2}}{q^2}\bigg)\bigg(1-\frac{a_{1}^{2}}{a^2}\bigg)+\bigg(1-\frac{q_{2}^{2}}{q^2}\bigg)\bigg(1-\frac{a_{2}^{2}}{a^2}\bigg) \notag\\
			&\quad +2\bigg(\frac{q_{2}q_{1}}{q^2}\bigg)\bigg(\frac{a_{2}a_{1}}{a^2}\bigg)\Bigg] 
			\end{align}
			This integral has can be solved numerically for specific choices of the spectral indices $\alpha$ and $\beta$.

		\subsection{Calculation of $\langle P(\textbf{k})_{\perp} \rangle_{\textbf{B}}$}
			\begin{align}
			\langle P(\textbf{k}_{\perp}) \rangle_{\textbf{B}}&=\int dx^3 \exp[i\textbf{k}_{\perp}\textbf{x}_{\perp}] \left(\partial_{1}^2(\textbf{x})-\partial_{2}^2(\textbf{x})+2i\partial_{1}(\textbf{x})\partial_{2}(\textbf{x})\right) |_{\textbf{J}} \exp\source \notag\\
			&=\int dx^3 \exp[i\textbf{k}_{\perp}\textbf{x}_{\perp}] \left(M_{11}^2(0)-M_{22}^2(0)+2i\Big(\frac{1}{2}M_{12}(0)+\frac{1}{2}M_{21}(0)\Big)\right) \notag\\
			&=0
			\end{align}
			The last step holds because of the relations $M_{11}(0)=M_{22}(0)$ and $M_{21}(0)=M_{12}(0)=0$.

		\subsection{Calculation of $\langle I(\textbf{k}_{\perp}) P(\textbf{k}'_{\perp}) \rangle_{\textbf{B}}$}
			\begin{align}
			&\langle I(\textbf{k}_{\perp}) P(\textbf{k}'_{\perp}) \rangle_{\textbf{B}}=\int dx^3 \int dx'^3 \exp \left[i(\textbf{k}_{\perp}\textbf{x}_{\perp}+\textbf{k}'_{\perp}\textbf{x}'_{\perp})\right] \Big(\partial_{1}^2(\fe{x})+\partial_{2}^2(\fe{x})\Big) \notag\\
			&\quad \ \Big(\partial_{1}^2(\fe{x}')-\partial_{2}^2(\fe{x}'+2i\partial_{1}(\fe{x}')\partial_{2}(\fe{x}'))\Big) \exp\source|_{\textbf{J}=0} \notag\\
			&=\int dx^3 \int dx'^3 \exp \big[i(\textbf{k}_{\perp}\textbf{x}_{\perp}+\textbf{k}'_{\perp}\textbf{x}'_{\perp})\big] \Big[M_{11}^2(0)-M_{22}^2(0)+2M_{11}^2(r)+2M_{22}^2(r) \notag\\
			&\quad \ +2M_{12}^2(r)-2M_{21}^2(r)+4iM_{21}(r)\Big(M_{11}(r)+M_{22}(r)\Big)\Big] \notag\\
			&=(2\pi)^{2} \delta^{2}(\fe{k}_{\perp}+\fe{k}_{\perp}') L_{z} \int dr^3 \int \frac{dq^3}{(2\pi)^3} \int \frac{dq'^3}{(2\pi)^3} \exp\big[i\fe{k}_{\perp}'\fe{r}_{\perp}\big] \exp\big[i\fe{r}(\fe{q}+\fe{q}')\big] \notag\\
			&\quad \ \Big[+2\hat{M}_{11}(q)\hat{M}_{11}(q')+2\hat{M}_{22}(q)\hat{M}_{22}(q')+2\hat{M}_{12}(q)\hat{M}_{12}(q')-2\hat{M}_{21}(q)\hat{M}_{21}(q')\notag\\
			&\quad \ +4i\hat{M}_{21}(q')\Big(\hat{M}_{11}(q)+\hat{M}_{22}(q)\Big)\Big] \notag\\
			&=(2\pi)^{2} \delta^{2}(\fe{k}_{\perp}+\fe{k}_{\perp}') L_{z} \int \frac{dq^3}{(2\pi)^3} \int \frac{dq'^3}{(2\pi)^3} \ (2\pi)^{3}\delta^{2}(\fe{q}_{\perp}+\fe{q}_{\perp}'+\fe{k}'_{\perp})\notag\\ 
			&\quad \ \delta(q_{z}+\tilde{q}_{z})\Big[2\hat{M}_{11}(q)\hat{M}_{11}(q')+2\hat{M}_{22}(q)\hat{M}_{22}(q')+2\hat{M}_{12}(q)\hat{M}_{12}(q')\notag\\
			&\quad \ -2\hat{M}_{21}(q)\hat{M}_{21}(q')+4i\hat{M}_{21}(q')\Big(\hat{M}_{11}(q)+\hat{M}_{22}(q)\Big)\Big] \notag\\
			&=(2\pi)^{2} \delta^{2}(\fe{k}_{\perp}+\fe{k}_{\perp}') L_{z} \int \frac{dq^3}{(2\pi)^3} \Big[2\hat{M}_{11}(q)\hat{M}_{11}(a)\notag\\
			&\quad \ +2\hat{M}_{22}(q)\hat{M}_{22}(a)+2\hat{M}_{12}(q)\hat{M}_{12}(a)-2\hat{M}_{21}(q)\hat{M}_{21}(a)\notag\\
			&\quad \ +4i\hat{M}_{21}(a)\Big(\hat{M}_{11}(q)+\hat{M}_{22}(q)\Big)\Big] \notag\\
			&=(2\pi)^{2} \delta^{2}(\fe{k}_{\perp}+\fe{k}_{\perp}') L_{z} \int \frac{dq^3}{(2\pi)^3} \bigg[\frac{\hat{M}_{N}(q) \hat{M}_{N}(a)}{q^2 a^2}\Big[2(q_{2}^2+q_{3}^2)(a_{2}^2+a_{3}^2)\notag\\
			&\quad \ +2(q_{1}^2+q_{3}^2)(a_{1}^2+a_{3}^2)-4i(q_{1}^2+q_{2}^2+2q_{3}^2)a_{2}a_{3}\Big]+2\frac{\hat{M}_{N}(q) \hat{H}(a)}{q^2 a} \notag\\
			&\quad \ \Big[iq_{1}q_{2}a_{3}-2(q_{1}^2+q_{2}^2+2q_{3}^2)a_{3}\Big]+\frac{\hat{M}_{N}(a) \hat{H}(q)}{q^2 a}\Big[ia_{1}a_{2}q_{3}\Big]\bigg] \notag\\
			&=(2\pi)^{2} \delta^{2}(\fe{k}_{\perp}+\fe{k}_{\perp}') L_{z} \int \frac{dq^3}{(2\pi)^3} \bigg[\frac{\hat{M}_{N}(q) \hat{M}_{N}(a)}{q^2 a^2}\Big[2(q_{2}^2+q_{3}^2)(a_{2}^2+a_{3}^2)\notag\\
			&\quad \ +2(q_{1}^2+q_{3}^2)(a_{1}^2+a_{3}^2)\Big]\bigg]
			\end{align}
			The last step was possible using an asymmetric property of the integrals. All terms in which $a_{3}$ occurs become zero when integrated from $-\infty$ to $\infty$. We can split up the integral into two parts covering the negative and positive regions and it can be seen that they will cancel each other. The remaining integral can be solved numerically for specific choices of the spectral indices $\alpha$ and $\beta$.

		\subsection{Calculation of $\langle I(\textbf{k}_{\perp}) \phi(\textbf{k}'_{\perp})\rangle_{\textbf{B}}$}
			\begin{align}
			\langle I(\textbf{k}_{\perp}) \phi(\textbf{k}'_{\perp})\rangle_{\textbf{B}}&=n_{e} a_{0} \int dx^3 \int dx'^3 \exp[i\textbf{k}_{\perp}\textbf{x}_{\perp}+i\textbf{k}'_{\perp}\textbf{x}'_{\perp}] \notag\\
			&\quad \left(\partial_{1}^2(\textbf{x}) \partial_{3}(\textbf{x}')+\partial_{2}^2(\textbf{x}) \partial_{3}(\textbf{x}')\right)|_{\textbf{J}=0} \exp\source \notag\\
			&=0 
			\end{align}

		\subsection{Calculation of $\langle P(\textbf{k}_{\perp}) \phi(\textbf{k}'_{\perp})\rangle_{\textbf{B}}$}
			\begin{align}
			\langle P(\textbf{k}_{\perp}) \phi(\textbf{k}'_{\perp})\rangle_{\textbf{B}}&=n_{e} a_{0} \int dx^3 \int dx'^3 \exp[i\textbf{k}_{\perp}\textbf{x}_{\perp}+i\textbf{k}'_{\perp}\textbf{x}'_{\perp}] \notag\\
			&\quad \left(\partial_{1}^2(\textbf{x}) \partial_{3}(\textbf{x}')-\partial_{2}^2(\textbf{x}) \partial_{3}(\textbf{x}')+2i\partial_{1}(\textbf{x})\partial_{2}(\textbf{x})\partial_{3}(\textbf{x}')\right)|_{\textbf{J}=0}\notag\\
			&\quad \exp\source \notag\\
			&=0 
			\end{align}

		\subsection{Calculation of $\langle I(\textbf{k}_{\perp}) \phi(\textbf{k}'_{\perp})\phi(\textbf{k}''_{\perp})\rangle_{\textbf{B}}$}
			\begin{align}
			&\langle I(\textbf{k}_{\perp}) \phi(\textbf{k}'_{\perp})\phi(\textbf{k}''_{\perp})\rangle_{\textbf{B}}=\int dx^3 \int dx'^3 \int dx''^3 \exp[i\textbf{k}_{\perp}\textbf{x}_{\perp}+i\textbf{k}'_{\perp}\textbf{x}'_{\perp}+i\textbf{k}''_{\perp}\textbf{x}''_{\perp}] \notag\\
			&\quad \left(\partial_{1}^{2}(\fe{x})\partial_{3}(\fe{x}')\partial_{3}(\fe{x}'')+\partial_{2}^{2}(\fe{x})\partial_{3}(\fe{x}')\partial_{3}(\fe{x}'')\right)|_{\fe{J}=0} \exp\source \notag\\
			&=\int dx^3 \int dx'^3 \int dx''^3 \exp[i\textbf{k}_{\perp}\textbf{x}_{\perp}+i\textbf{k}'_{\perp}\textbf{x}'_{\perp}+i\textbf{k}''_{\perp}\textbf{x}''_{\perp}] \Big[\underbrace{M_{11}(0)M_{33}(\fe{x}''-\fe{x}')}_{\mathrm{Part1}} \notag\\
			&\quad+\underbrace{2\left(M_{31}(\fe{x}''-\fe{x})M_{31}(\fe{x}'-\fe{x})\right)}_{\mathrm{Part 2}}+\underbrace{M_{22}(0)M_{33}(\fe{x}''-\fe{x}')}_{\mathrm{Part 3}} \notag\\
			&\quad+\underbrace{2\left(M_{32}(\fe{x}''-\fe{x})M_{32}(\fe{x}'-\fe{x})\right)}_{\mathrm{Part 4}}\Big]
			\end{align}
			\textbf{Part 1 and 3} can be calculated in the same way, as they only differ in the component of $\fe{M}$ chosen by the derivatives ($i=1,2$):
			\begin{align}
			&\int dx^3 \int dx'^3 \int dx''^3 \exp[i\textbf{k}_{\perp}\textbf{x}_{\perp}+i\textbf{k}'_{\perp}\textbf{x}'_{\perp}+i\textbf{k}''_{\perp}\textbf{x}''_{\perp}] M_{ii}(0)M_{33}(\fe{x}''-\fe{x}') \notag\\
			&\overset{\rlap{\textbf{x}''=\textbf{x}'+\textbf{r}}}{=}\qquad M_{ii}(0) \int dx^3 \int dx'^3 \int dr^{3} M_{33}(\fe{r}) \exp[i\fe{x}'_{\perp}(\fe{k}'_{\perp}+\fe{k}''_{\perp})] \exp[i\fe{r}_{\perp}\fe{k}''_{\perp}] \notag\\ 
			&\quad \exp[i\fe{k}_{\perp}\fe{x}_{\perp}] \notag\\
			&=M_{ii}(0) L_{z}^{2} (2\pi)^{4} \delta^{2}(\fe{k}'_{\perp}+\fe{k}''_{\perp}) \delta^{2}(\fe{k}_{\perp})\int d^{3}r M_{33}(\fe{r}) \exp[i\fe{r}_{\perp}\fe{k}''_{\perp}] \notag\\
			&=M_{ii}(0) L_{z}^{2} (2\pi)^{4} \delta^{2}(\fe{k}'_{\perp}+\fe{k}''_{\perp}) \delta^{2}(\fe{k}_{\perp})\int d^{3}r \int \frac{dq^{3}}{(2\pi)^{3}} \hat{M}_{33}(\fe{q}) \notag\\
			&\quad \exp[i\fe{r}_{\perp}(\fe{k}''_{\perp}-\fe{q}_{\perp})] \exp[ir_{z}q_{z}] \notag\\
			&=M_{ii}(0) L_{z}^{2} (2\pi)^{4} \delta^{2}(\fe{k}'_{\perp}+\fe{k}''_{\perp}) \delta^{2}(\fe{k}_{\perp}) \int dq^{3} \hat{M}_{33}(\fe{q}) \delta^{2}(\fe{k}''_{\perp}-\fe{q}_{\perp}) \delta(-q_{z}) \notag\\
			&=M_{ii}(0) L_{z}^{2} (2\pi)^{4} \delta^{2}(\fe{k}'_{\perp}+\fe{k}''_{\perp}) \delta^{2}(\fe{k}_{\perp}) \hat{M}_{33}(\fe{k}''_{\perp},0)
			\end{align}
			Equally, \textbf{Part 2} and \textbf{Part 4} can be solved on the same basis ($i=1,2$):
			\begin{align}
			&2\int dx^3 \int dx'^3 \int dx''^3 \exp[i\textbf{k}_{\perp}\textbf{x}_{\perp}+i\textbf{k}'_{\perp}\textbf{x}'_{\perp}+i\textbf{k}''_{\perp}\textbf{x}''_{\perp}] M_{3i}(\fe{x}''-\fe{x}) \notag\\
			&\quad M_{3i}(\fe{x}'-\fe{x}) \notag\\
			&=2\int dx^3 \int dr^3 \int dr'^3 \exp[i\fe{x}_{\perp}(\fe{k}_{\perp}+\fe{k}'_{\perp}+\fe{k}''_{\perp})] \exp[i\fe{r}'_{\perp}\fe{k}''_{\perp}] \exp[i\fe{r}_{\perp}\fe{k}'_{\perp}] \notag\\
			&\quad M_{3i}(\fe{r}') M_{3i}(\fe{r}) \notag\\
			&=2 L_{z} (2\pi)^2 \delta^{2}(\fe{k}_{\perp}+\fe{k}'_{\perp}+\fe{k}''_{\perp}) \int dr^3 \int dr'^3 \int \frac{dq^{3}}{(2\pi)^{3}} \int \frac{dq'^{3}}{(2\pi)^{3}} \exp[i\fe{r}_{\perp}'(\fe{k}_{\perp}''-\fe{q}_{\perp}')] \notag\\
			&\quad \exp[-ir_{z}'q_{z}'] \exp[i\fe{r}_{\perp}(\fe{k}_{\perp}'-\fe{q}_{\perp})] \exp[-ir_{z}q_{z}] \hat{M}_{3i}(\fe{q}')) \hat{M}_{3i}(\fe{q}) \notag\\
			&=2 L_{z} (2\pi)^2 \delta^{2}(\fe{k}_{\perp}+\fe{k}'_{\perp}+\fe{k}''_{\perp}) \hat{M}_{3i}(\fe{k}_{\perp}'',0) \hat{M}_{3i}(\fe{k}_{\perp}',0)
			\end{align}
			This gives as final result:
			\begin{align}
			&\langle I(\textbf{k}_{\perp}) \phi(\textbf{k}'_{\perp})\phi(\textbf{k}''_{\perp})\rangle_{\textbf{B}}=L^{2} (2\pi)^{4} \delta^{2}(\fe{k}'_{\perp}+\fe{k}''_{\perp}) \delta^{2}(\fe{k}_{\perp}) \hat{M}_{33}(\fe{k}''_{\perp},0) \notag\\  
			&\quad \big[M_{11}(0)+M_{22}(0)\big]+2n_{e}^{2} L_{z} (2\pi)^2 \delta^{2}(\fe{k}_{\perp}+\fe{k}'_{\perp}+\fe{k}''_{\perp})\big[\hat{M}_{31}(\fe{k}_{\perp}'',0) \hat{M}_{31}(\fe{k}_{\perp}',0) \notag\\
			&\quad +\hat{M}_{32}(\fe{k}_{\perp}'',0) \hat{M}_{32}(\fe{k}_{\perp}',0)\big]\notag\\
			&=L^{2} (2\pi)^{4} \delta^{2}(\fe{k}'_{\perp}+\fe{k}''_{\perp}) \delta^{2}(\fe{k}_{\perp})\hat{M}_{N}(u)2M_{N}(0) \notag\\
			&\quad-2n_{e}^{2} L_{z} (2\pi)^2 \delta^{2}(\fe{k}_{\perp}+\fe{k}'_{\perp}+\fe{k}''_{\perp}) \frac{\hat{H}(u)\hat{H}(v)}{uv}\Big(u_{1}v_{1}+u_{2}v_{2}\Big) \label{al:Ipp} 
			\end{align}
			A discussion and a plot of this important result can be found in Sec. \ref{other}.

		\subsection{Calculation of $\langle P(\textbf{k}_{\perp}) \phi(\textbf{k}'_{\perp})\phi(\textbf{k}''_{\perp})\rangle_{\textbf{B}}$}
			\begin{align}
			&\langle P(\textbf{k}_{\perp}) \phi(\textbf{k}'_{\perp})\phi(\textbf{k}''_{\perp})\rangle_{\textbf{B}}=\int dx^3 \int dx'^3 \int dx''^3 \exp[i\textbf{k}_{\perp}\textbf{x}_{\perp}+i\textbf{k}'_{\perp}\textbf{x}'_{\perp}+i\textbf{k}''_{\perp}\textbf{x}''_{\perp}] \notag\\
			&\quad \left(\partial_{1}^2(\fe{x})\partial_{3}(\fe{x}')\partial_{3}(\fe{x}'')-\partial_{2}^2(\fe{x})\partial_{3}(\fe{x}')\partial_{3}(\fe{x}'')+2i\partial_{1}(\fe{x})\partial_{2}(\fe{x})\partial_{3}(\fe{x}')\partial_{3}(\fe{x}'')\right)|_{\fe{J}=0} \notag\\ 
			&\quad \exp\source \notag\\
			&=\int dx^3 \int dx'^3 \int dx''^3 \exp[i\textbf{k}_{\perp}\textbf{x}_{\perp}+i\textbf{k}'_{\perp}\textbf{x}'_{\perp}+i\textbf{k}''_{\perp}\textbf{x}''_{\perp}] \notag\\
			&\quad \Big[M_{33}(\fe{x}''-\fe{x}')\Big(M_{11}(0)+M_{22}(0)\Big)+2M_{31}(\fe{x}''-\fe{x})M_{31}(\fe{x}'-\fe{x}) \notag\\
			&\quad+2M_{32}(\fe{x}''-\fe{x})M_{32}(\fe{x}'-\fe{x})+2iM_{32}(\fe{x}''-\fe{x})M_{31}(\fe{x}'-\fe{x}) \notag\\
			&\quad+2iM_{32}(\fe{x}'-\fe{x})M_{31}(\fe{x}''-\fe{x})\Big]
			\end{align}
			This can be evaluated in exactly the same way as (\ref{al:Ipp}) because all the terms have the same basic structure. This provides:
			\begin{align}
			&\langle P(\textbf{k}_{\perp}) \phi(\textbf{k}'_{\perp})\phi(\textbf{k}''_{\perp})\rangle_{\textbf{B}}=L^{2} (2\pi)^{4} \delta^{2}(\fe{k}'_{\perp}+\fe{k}''_{\perp}) \delta^{2}(\fe{k}_{\perp})\hat{M}_{33}(\fe{k}''_{\perp},0)[M_{11}(0) \notag\\
			&\quad+M_{22}(0)]+2n_{e}^{2} L_{z} (2\pi)^2 \delta^{2}(\fe{k}_{\perp}+\fe{k}'_{\perp}+\fe{k}''_{\perp}) \notag\\
			&\quad \Big[\hat{M}_{31}(\fe{k}_{\perp}'',0) \hat{M}_{31}(\fe{k}_{\perp}',0)+\hat{M}_{32}(\fe{k}_{\perp}'',0) \hat{M}_{32}(\fe{k}_{\perp}',0)+i\hat{M}_{32}(\fe{k}_{\perp}'',0) \hat{M}_{31}(\fe{k}_{\perp}',0) \notag\\
			&\quad+i\hat{M}_{31}(\fe{k}_{\perp}'',0) \hat{M}_{32}(\fe{k}_{\perp}',0)\Big] \notag\\
			&=+2n_{e}^{2} L_{z} (2\pi)^2 \delta^{2}(\fe{k}_{\perp}+\fe{k}'_{\perp}+\fe{k}''_{\perp})  \frac{\hat{H}(u)\hat{H}(v)}{uv}  \Big(u_{1}v_{1}-u_{2}v_{2} \ +i\big(u_{1}v_{2}+u_{2}v_{1}\big)\Big)
			\end{align}
			A discussion and a plot of this important result can be found in Sec. \ref{other}.

\section[Synchrotron radiation]{Radio observables, synchrotron radiation and Stokes parameters}\label{sync}
	This section introduces the notation to describe radio observables. For our statistical approach, synchrotron radiation is the fundamental observed quantity on which our deduction is based. Since we are attempting to infer properties of the magnetic field statistics through statistics of the radio synchrotron observables, we need a clear and compact notation of these observables.

	All accelerated charges emit electromagnetic radiation. If accelerated by a magnetic field, the radiation is called \textit{cyclotron radiation} in case of nonrelativistic and \textit{synchrotron radiation} in case of relativistic velocities. With regard to astrophysics the latter is far more important. This is because of the much higher power radiated by relativistic particles, since the total emitted power of an accelereated charge depends on $\gamma^2$, its Lorentz factor squared \citep{1979rpa..book.....R}:
	\begin{align}
	P=\frac{4}{3}\sigma_{T}c\beta^2\gamma^2\frac{B^2}{8\pi},
	\end{align}
	where $\beta=v/c$.
	Synchrotron radiation has a characteristic polarisation, with a high percentage of linear polarisation. Furthermore, the relativistic beaming effect confines the energy radiated within a cone around the direction of the moving charge. Due to these features and since a large fraction of astrophysical synchrotron emission falls into radio wavebands, it is relatively easy to detect and provides us with an excellent way to observe and study magnetic fields. 
	
	Synchrotron radiation is mainly emitted by relativistic electrons. Other charged particles like protons contribute far less to the radiated power due to their larger mass.
	Following \citet{1979rpa..book.....R} we assume a power-law distribution of the cosmic ray electron energies with spectral index $p$ 
	\begin{align}
	N(\gamma) \ d\gamma=C\gamma^{-p} d\gamma,
	\end{align}
	where $C$ is a normalisation factor which determines the number density of relativistic electrons.
	The total power emitted per unit volume and unit frequency by such distributed electrons, assumed to have an isotropic pitch-angle distribution, is then given by the integral over $N(\gamma) \ d\gamma$ times the single particle radiation spectrum. It can be shown that this leads to a power law in synchrotron emmissivity \citep{1979rpa..book.....R}:
	\begin{align}
	j \propto \omega^{-\frac{(p-1)}{2}} B_{\perp}^{\frac{(p+1)}{2}} C .
	\end{align}
	In the end we are interested in observable quantities, namely the total and polarised intensity of the observed region.
	In terms of emissivity $j$ they are:
	\begin{align}
	&\mathrm{total \ intensity} &&I(\fe{x}_{\perp})=\int dz \ j(\fe{x}) , \ \mathrm{and} \\
	&\mathrm{polarised \ intensity} &&P(\fe{x}_{\perp})=\int dz \ j(\fe{x}) f(p) \exp(2i\chi(\fe{x})),\label{poli}
	\end{align}
	where $f(p)=(p+1)/(p+7/3)$ is the polarisation fraction and the integrals are along the line of sight from the source to the observer. The angle $\chi$ is the polarisation angle of the radiation. With $\chi$ we can introduce the effect of Faraday-rotation into our formulas. 
	Faraday rotation is the rotation of the polarisation plane of a linearly polarised wave in a medium with a non-scalar dielectric constant due to a magnetic field. In such an environment, the dielectric constant differs for left and right circular polarisation \citep{1979rpa..book.....R}:
	\begin{equation}
	\epsilon=1-\frac{\omega_{p}^2}{\omega(\omega \pm \omega_{B})},
	\end{equation}
	where $\omega$ denotes the frequency of the wave and $\omega_{p}=eB/mc$ is the cyclotron frequency. Plus and minus signs denote the case for right and left circular polarisation respectively. If we describe a linearly polarised wave as a superposition of a wave with right circular polarisation and a wave with left circular polarisation, the linear polarisation plane will not remain constant. 
	
	Including Faraday rotation, the total angle $\chi$ at the location of the observer is given by
	\begin{align}
	\chi(\fe{x})=\chi_{0}(\fe{x})+\lambda^2\phi(\fe{x}).
	\end{align}
	Here $\chi_{0}$ denotes the polarisation angle at the position of emission whereas the Faraday depth $\phi(\fe{x})$ is defined as \citep{2008ApJ...676...70K} 
	\begin{align}
	\phi(\fe{x})=\frac{e^3}{2\pi \ n_{e}^2c^4} \int dz \ n_{e} \ B_{3} .
	\end{align}
	It describes the phase angle through which the electric vector rotates due to Faraday rotation. For this study, it is assumed that the electron density  $n_{e}$ is constant in order to simplify the calculations. Furthermore, we restrict ourselves to observations at short wavelength $\lambda$, thus ignoring the term $\exp(2i\lambda^2 \phi(\fe{x}))$ in (\ref{poli}), which would lead to so called intrinsic Faraday-rotation. Nevertheless $\phi$ is used as an independent radio observable, allowing us to infer information about the component of the magnetic field parallel to the line of sight. In real observations this would be possible by finding a background source to probe $\phi$ in the observed region (see Fig. \ref{obsreg}). Sometimes we refer to the notion rotation measure (RM) instead of Faraday depth. Correctly this describes the factor between the rotation angle and $\lambda^2$ obtained through observations. However, we use it in the context of real observations, because it is the more conventional term than Faraday depth when dealing with data.
		
	$I(\fe{x}_{\perp})$ and $P(\fe{x}_{\perp})$ can be expressed most suitable using the Stokes parameters. After some calculations 
	\citep[see e.g.][]{1979rpa..book.....R,2009MNRAS.398.1970W} the first three Stokes parameters $I$,$Q$ and $U$ can be determined for synchrotron radiation of a power-law spectrum distributed, isotropic, relativistic electron population :
	\begin{align}
	&I=2\ F(p) \ \omega^{\frac{(1-p)}{2}} \int dz \ \big(B_{1}^2+B_{2}^2\big)^{\frac{(p-3)}{4}} \big(B_{1}^2+B_{2}^2\big) , \\
	&Q=2 \ G(p) \ \omega^{\frac{(1-p)}{2}} \int dz \ \big(B_{1}^2+B_{2}^2\big)^{\frac{(p-3)}{4}} \big(B_{1}^2-B_{2}^2\big) , \\
	&U=2 \ G(p) \  \omega^{\frac{(1-p)}{2}} \int dz \ \big(B_{1}^2+B_{2}^2\big)^{\frac{(p-3)}{4}} 2B_{1}B_{2} . 
	\end{align}
	The two functions $F(p)$ and $G(p)$ are expressed in terms of physical constants and gamma functions of the spectral index $p$ \citep[see][]{2009MNRAS.398.1970W,1998A&A...330...90E}:
	\begin{align}
	&F(p)=\frac{\sqrt[3]{3} \ e^3}{32 \pi^2 m_{e} c^2} \big(\frac{2m_{e}c}{3e}\big)^{\frac{(1-p)}{2}} C \ \Gamma\big(\frac{p}{4}-\frac{1}{12}\big)\Gamma\big(\frac{p}{4}+\frac{19}{12}\big) \frac{2^{(p+1)}{2}}{p+1} , \\
	&G(p)=\frac{\sqrt[3]{3} \ e^3}{32 \pi^2 m_{e} c^2} \big(\frac{2m_{e}c}{3e}\big)^{\frac{(1-p)}{2}} C \ \Gamma\big(\frac{p}{4}-\frac{1}{12}\big)\Gamma\big(\frac{p}{4}+\frac{7}{12}\big) 2^{(p-3)}{2} . 
	\end{align}
	Throughout this study, the value $p=3$ has been adopted for the spectral index of the electron distribution. This not only simplifies calculations, it is also a reasonable choice from a physical point of view. Typical values for the spectral index of relativistic electrons in our galaxy measured directly by cosmic rays on Earth or indirectly via their synchrotron emission are around $p\approx2.7$ \citep{Amsler:1124989}.
	Deviations from $p=3$ can be added to the results of this work later in terms of corrections.
	
	Finally we can state $I$, $P$ and $\phi$ in the form to be used in this study.  
	During our calculations, all fore factors will be suppressed for convenience, resulting in
	\begin{align}
	&I=\int dz \big(B_{1}^2+B_{2}^2\big), \\
	&P=\int dz \big(B_{1}^2-B_{2}^2+2iB_{1}B_{2}\big),\quad \text{and} \\
	&\phi=\int dz B_{3}
	\end{align}

\section[Covariance matrix and correlation tensor]{Identity of the covariance matrix with the correlation tensor for Gaussian statistics}\label{cocor}
	In Sec. \ref{sec:gen} we use the fact, that the covariance matrix of a Gaussian probability distribution of the magnetic field is identical to the magnetic correlation tensor. 
	From the Gaussian probability distribution
	$$\mathcal{G}(B,\tilde{M})=\frac{1}{\sqrt{|2\pi\tilde{M}|}}\exp[-\frac{1}{2}B^{+}\tilde{M}^{-1}B]$$
	the identity is easily to shown:
	\begin{align}
	M&=\langle BB^{\dagger}\rangle=\int \mathcal{D}B \  \mathcal{G}(B,\tilde{M}) B B^{\dagger} \notag\\
	&=\Big(\frac{\partial}{\partial J}\frac{\partial}{\partial J^{+}}\Big)\Big|_{J=0}\int \mathcal{D}B \frac{1}{\sqrt{|2\pi \tilde{M}|}} \exp[-\frac{1}{2}B^{\dagger}\tilde{M}^{-1}B+J^{\dagger}B] \notag\\
	&=\Big(\frac{\partial}{\partial J}\frac{\partial}{\partial J^{\dagger}}\Big)\Big|_{J=0}\exp[\frac{1}{2}J^{\dagger}\tilde{M}J] \notag\\
	&=([\tilde{M}+\tilde{M}J]\exp[\frac{1}{2}J^{+}\tilde{M}J)]|_{J=0}=\tilde{M} \notag
	\end{align}
	In this case, we have used the generating-function technique which is explained in \ref{sec:gen}.

\end{appendix}

\end{document}